\documentclass[12pt,preprint]{aastex}

\usepackage{color}

\begin{document}


\title{INITIATION OF CME AND ASSOCIATED FLARE CAUSED BY HELICAL KINK INSTABILITY OBSERVED BY SDO/AIA}

%
%
\author{PANKAJ KUMAR\altaffilmark{1}, K.-S. CHO\altaffilmark{2,3,1}, S.-C. BONG\altaffilmark{1}, SUNG-HONG PARK\altaffilmark{1}, Y.H. KIM\altaffilmark{1}}
\altaffiltext{1}{Korea Astronomy and Space Science Institute (KASI), Daejeon, 305-348, Republic of Korea}
\altaffiltext{2}{NASA Goddard Space Flight Center, Greenbelt, Maryland,
USA.}
\altaffiltext{3}{Department of Physics, The Catholic University of America, Washington, D. C., USA.}
\email{pankaj@kasi.re.kr}

%
\begin{abstract}
In this paper, we present multiwavelength observations of helical kink instability as a trigger of a CME which occurred in AR NOAA 11163 on 24 February 2011. The CME was associated with a M3.5 limb flare. High resolution observations from SDO/AIA suggest the development of helical kink instability in the erupting prominence, which implies a flux rope structure of the magnetic field. A brightening starts below the apex of the prominence with its slow rising motion ($\sim$100 km s$^{-1}$) during the activation phase.  A bright structure, indicative of a helix with $\sim$3-4 turns, was transiently formed at this position. The corresponding twist of $\sim$6$\pi$-8$\pi$ is sufficient to generate the helical kink instability in a flux rope according to recently developed models. A slowly rising blob structure was subsequently formed at the apex of the prominence, and a flaring loop was observed near the footpoints. Within two minutes, a second blob was formed in the northern prominence leg. The second blob erupts (like a plasmoid ejection) with the detachment of the northern prominence leg, and flare intensity maximizes. The first blob at the prominence apex shows rotational motion in the counterclockwise direction in the plane of sky, interpreted as unwinding motion of a helix, and it also erupts to give the coronal mass ejection (CME). RHESSI hard X-ray sources show the two footpoint sources and a loop-top source during the flare. We found RHESSI hard X-ray flux, soft X-ray flux derivative and CME acceleration in the low corona correlate well, which is in agreement with the standard flare model (CSHKP). We also discuss the possible role of ballooning as well as torus instabilities in driving the CME. We conclude that the CME and flare were triggered  by the helical kink instability in a flux rope  and accelerated mainly by the torus instability.
 
 \end{abstract}
\keywords{Solar flare -- coronal loops, magnetic field, sunspots, coronal mass ejection, magnetic reconnection}

\section{INTRODUCTION}
Coronal Mass Ejections (CMEs) are the expulsion of plasma and magnetic fields from the Sun to the interplanetary space \citep{hund1984,sch1996}. The reconnection of twisted magnetic fields is widely considered to be potentially important for the triggering of solar eruptive phenomena, i.e. flares and CMEs.   
Sigmoid structures in the corona indicate the presence of twisted magnetic structures and the active regions containing `S' shaped or `inverse S' shaped sigmoids have greater tendency to erupt \citep{canfield1999}. Filament/flux rope interactions caused by sunspot rotation may cause to initiate the eruption \citep{kumar2010a,kumar2010b,chandra2011,torok2011}. Several theoretical models of CMEs initiation have been proposed. For example, the flux cancellation model
describes the formation of a flux rope and its subsequent
destabilization by the weakening of overlying flux \citep{amari2000,amari2003a,amari2003b,linker2003, roussev2004}. Breakout model requires strongly sheared magnetic field along the neutral line and the quadrupolar field structure along with the presence of magnetic null \citep{antiochos1998,antiochos1999}. The sheared central arcade causes the field lines to rise and slow reconnection at the null point transfers overlying flux in the central arcade to the neighboring arcades, which destabilizes the central arcade (e.g., \citealt{joshi2007}).
The kink instability of twisted magnetic flux tubes in the solar corona is considered as a possible candidate to initiate solar eruptions  \citep{torok2003,torok2004,kliem2004,cho2009,kliem2010}. It develops in line-tied magnetic fields if the twist exceeds a threshold value of $2.5\pi$ \citep{hood1981,einaudi1983}. Recent numerical MHD modeling indicates that a twist of $\sim3.5\pi$, equivalent to 1.75 field line turns, is a typical threshold value under coronal conditions \citep{fan2003,fan2004,kliem2004,torok2003,torok2004}. Helical flux tube with multiple turns (3-4 turns) associated with double flare and CME events has been reported by \citet{gary2004} and \citet{liu2003}. Recently, \citet{sri2010} and \citet{kumar2010c} have also reported the activation of helical kink instability in the flux rope which caused flares associated with failed eruption. The helical structure showed $\sim$2-3 turns, which was well above the critical twist needed for kink instability. 

 \citet{kliem2006} studied
 the stability of a toroidal current ring immersed in a background
potential magnetic field by extending the results of \citet{bateman1978}.
\citet{bateman1978} derived the threshold condition for this instability given by 

\begin{equation}
 n=-\frac{d (log B)}{d (log R)}>1.5
 \end{equation}

Where B denotes the background magnetic field strength 
at a geometrical height R above the eruption region.
This instability is referred as torus instability \citep{kliem2006}. If the decay index (n)
of the overlying magnetic field (B) approaches critical value, it results in torus instability 
or partial torus instability \citep{torok2005,olmedo2010}. This instability has been suggested 
to be relevant to drive a CME and discussed by several authors \citep{torok2007,fan2007,aul2010,dem2010}.

Moreover, high-beta (both thermal and kinetic) plasmas may be considered ready for eruption. 
The high-beta condition does not necessarily require high temperature, because the gas 
density in the prominences/chromosphere is large enough to compensate lower temperature \citep{shibasaki2001}.
\citet{shibasaki1999} showed the formation of isolated radio sources over the loop top 
of a flaring loop at several times during the flare development and suggested that 
these isolated loop-top radio sources are magnetic islands or balloons produced by
 nonlinear ballooning instabilities due to the high-beta plasma in the loop. Furthermore, \citet{sakai1982} has shown by using a drift analogy between the centrifugal and gravitational forces that the origin of a CME could be attributed to the ballooning instability excited by fast magnetosonic waves generated by a flare energy release outside the loop.

Now, high resolution observations from SDO/AIA are providing the opportunity to study the detailed morphological changes in the flux rope during the development of helical kink instability for CME initiation.  We present the rare multiwavelength observations of helical kink instability of the flux rope using high resolution observations from SDO/AIA. We have used the RHESSI hard X-ray observations to see the evolution of hard X-ray sources during the CME initiation associated with M-class flare. In section 2, we present the multiwavelength observations and data analysis. In Section 3, we present the theoretical estimations to test ballooning and torus instabilities in the flux rope and in the last section we discuss our results and conclusion. 

%
%
\section{OBSERVATIONS AND DATA}
The active region (AR) NOAA 11163 was lying near the eastern limb ($\sim$N19E77) on 24 February 2011. An M3.5 flare was occurred in this AR, which started at 07:23 UT, peaked at 07:35 UT, and ended at 07:42 UT, according to the NOAA event reports (refer to Figure \ref{flux}).  The flare was very impulsive in nature. SDO/AIA observes multiwavelength images at 0.6$\arcsec$ per pixel resolution with 12 s cadence. The field of view is 1.3 R$_\odot$. To investigate the CME initiation and associated flare, we have used AIA 171 \AA \ (Fe IX, T$\sim$0.6 MK), 131 \AA \ (Fe XX/XXIII, T$\sim$11 MK), 304 \AA \ (He II, T$\sim$0.05 MK) and 1600 \AA \ (C IV+ continuum, T$\sim$0.1 MK) images from chromospheric to coronal heights. We have used the SDO/HMI data for the morphology of the active region. 

\begin{figure*}
\centering{
\includegraphics[width=5.5cm]{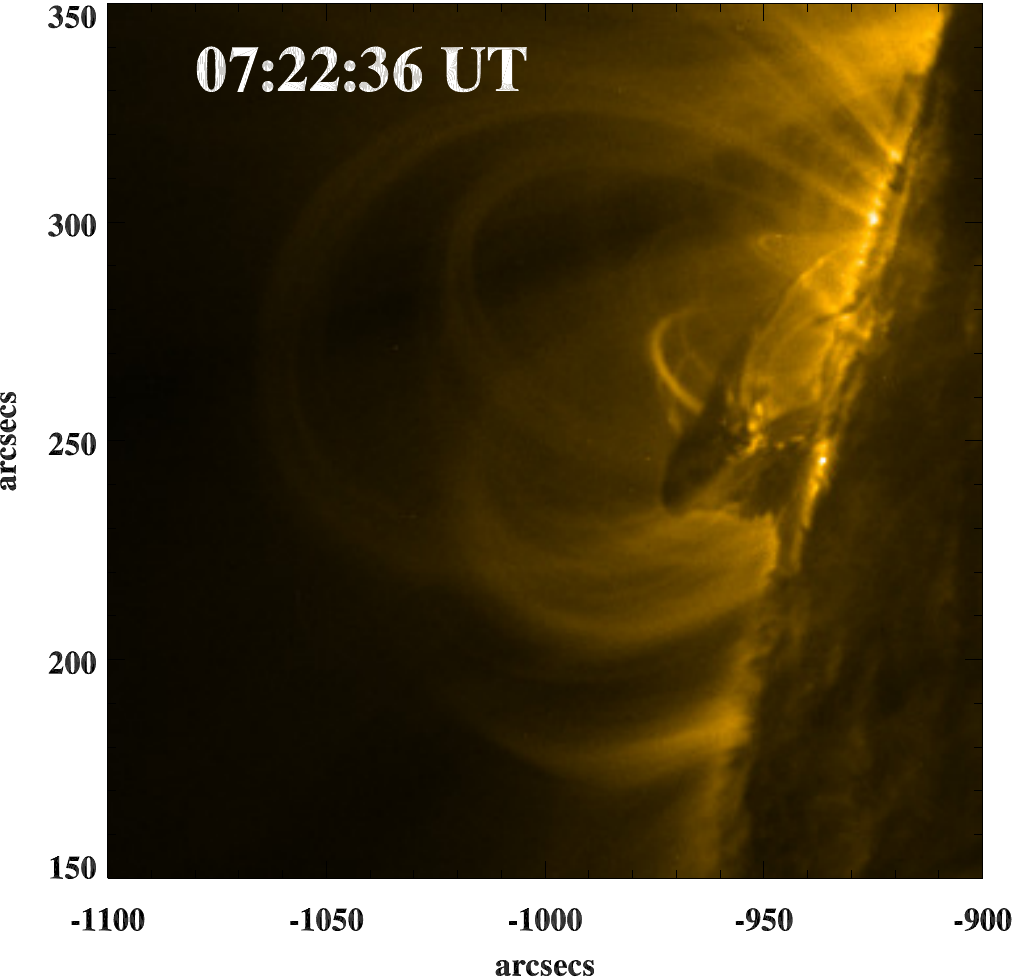}
\includegraphics[width=5.5cm]{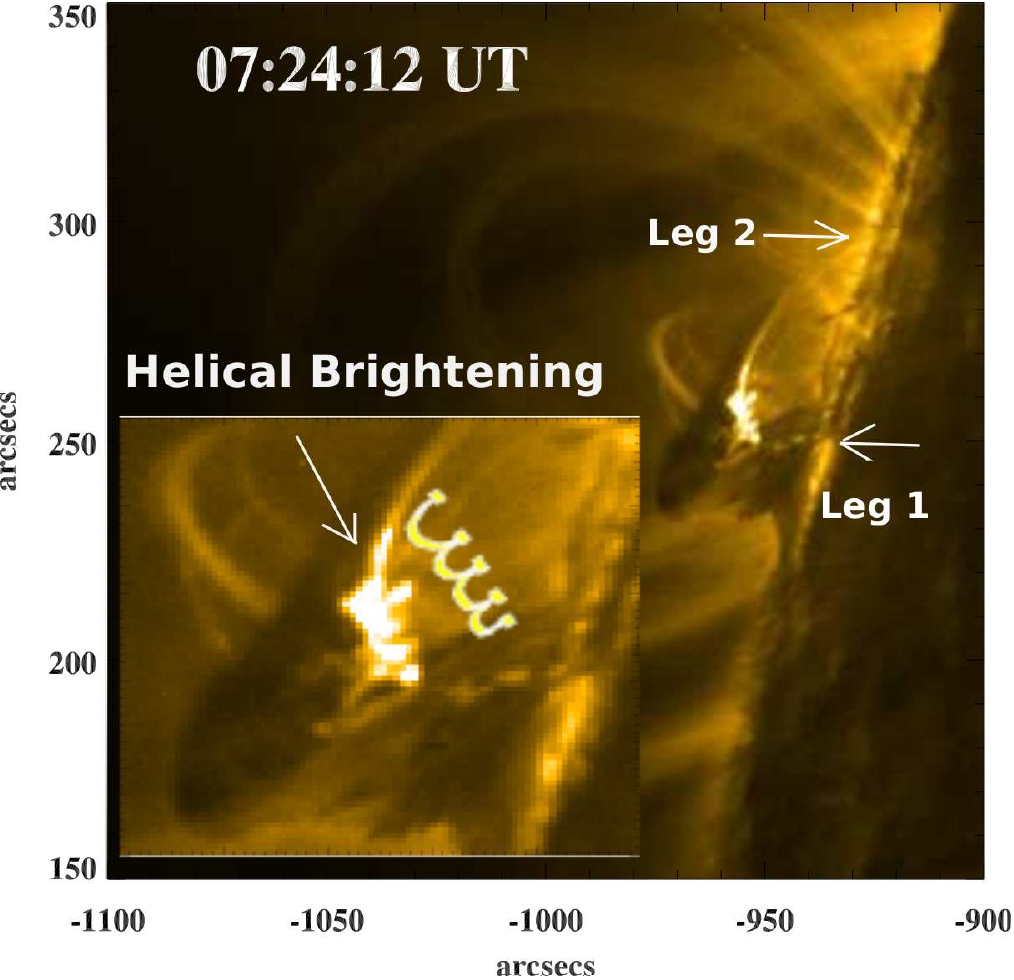}
}
\centering{
\includegraphics[width=5.5cm]{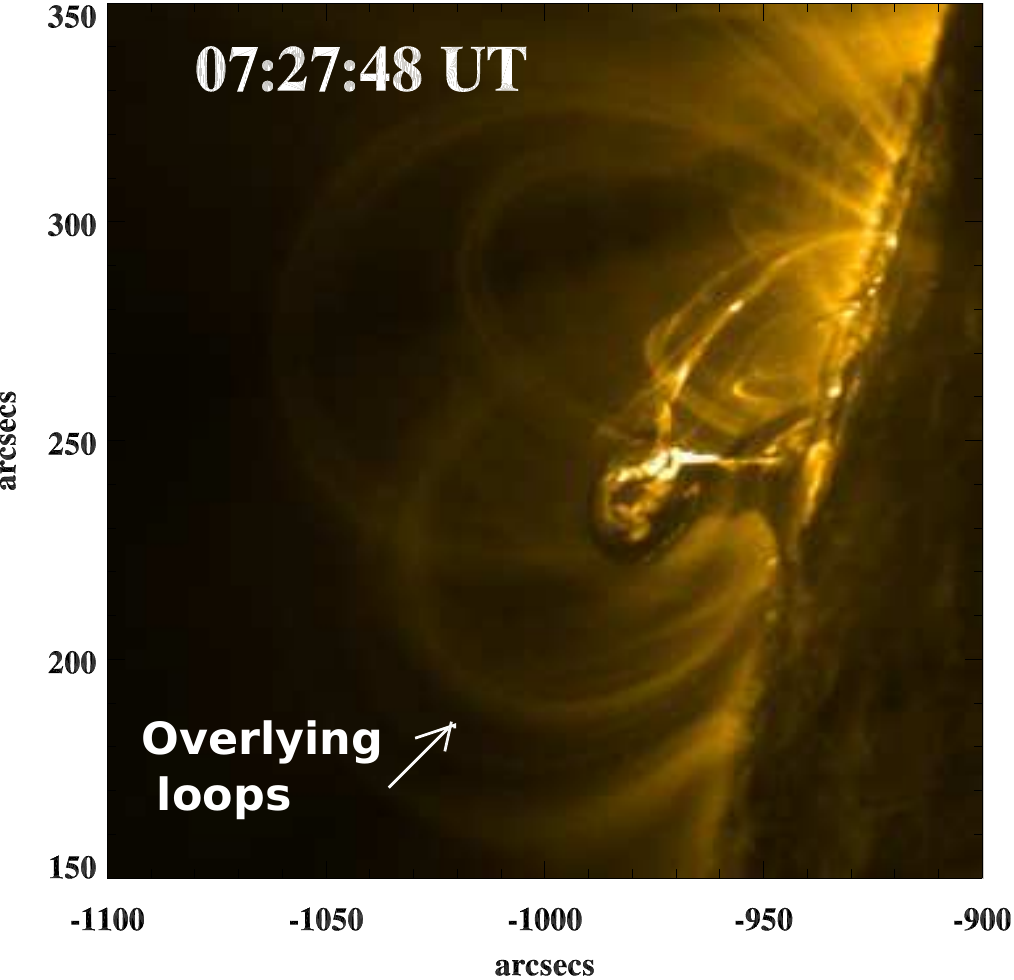}
\includegraphics[width=5.5cm]{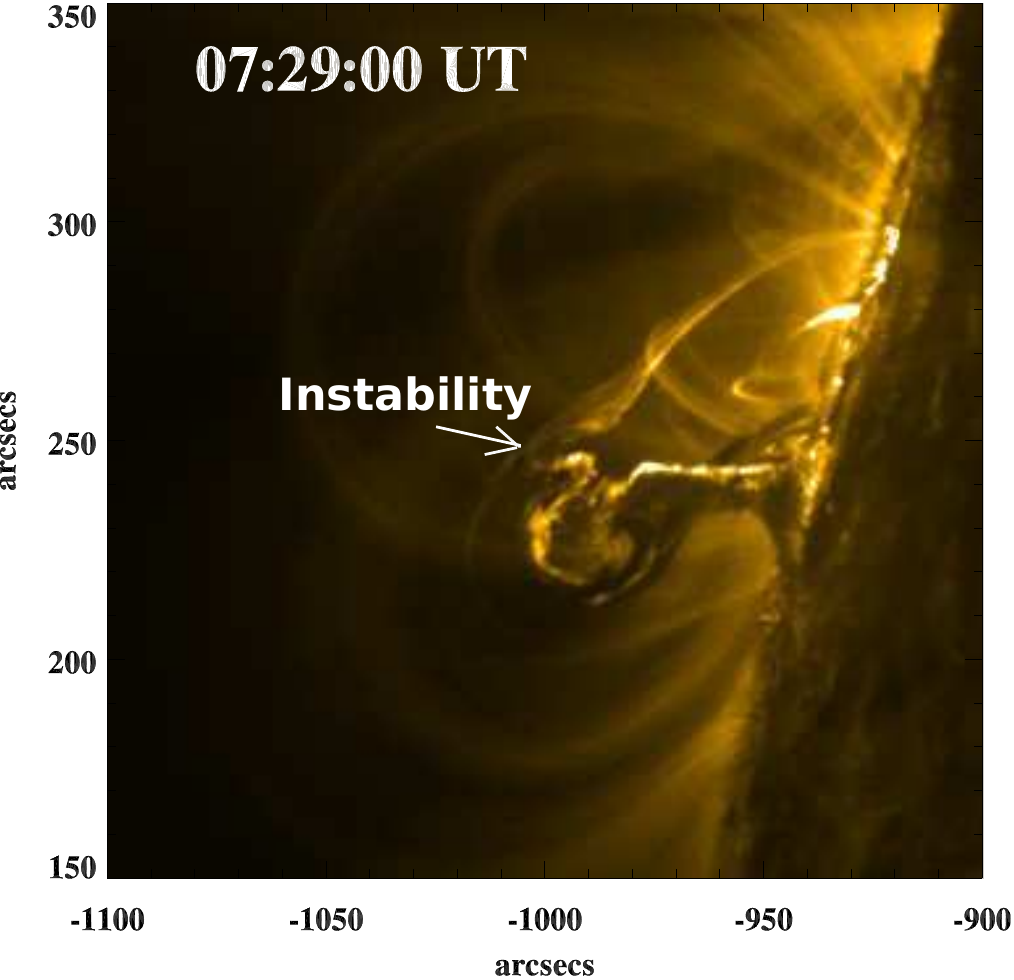}
}
\centering{
\includegraphics[width=5.5cm]{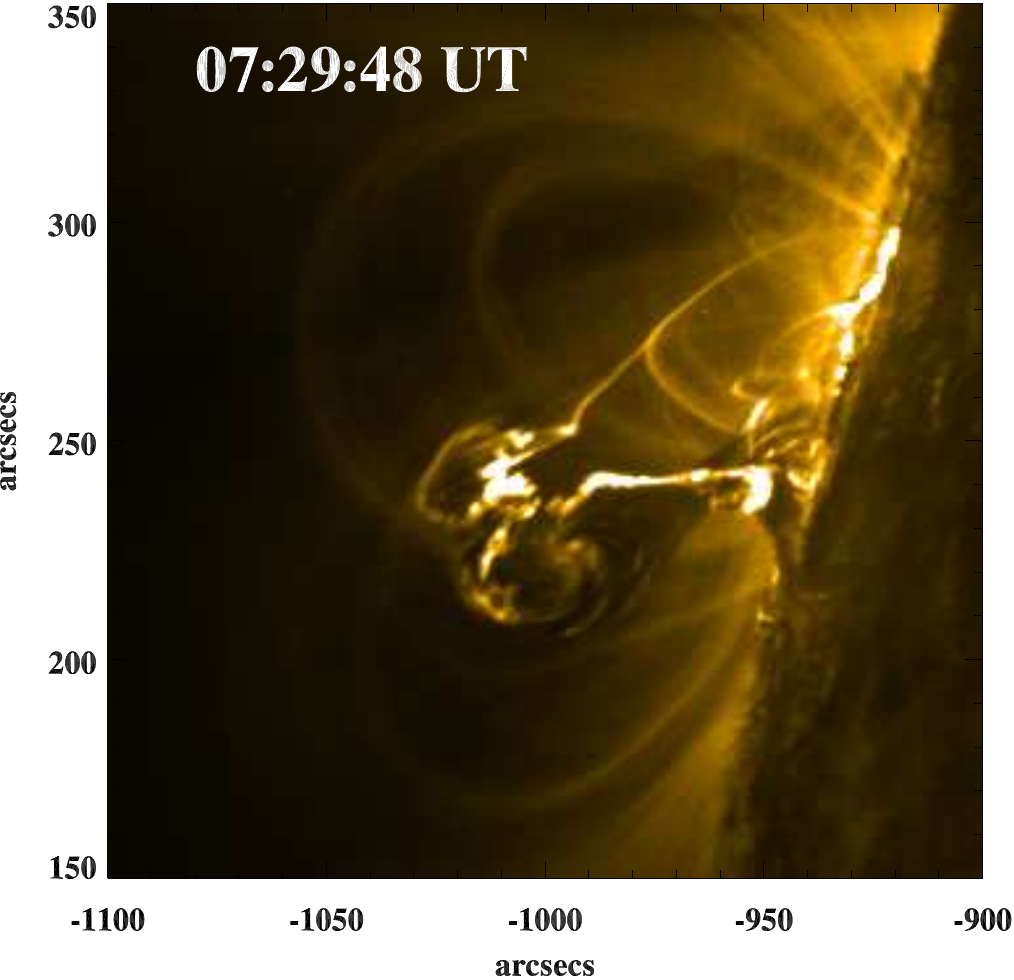}
\thicklines
$ \color{white} \put(-110,60){\vector(1,0){15}} \color{white} \put(-120,57){A}$
$ \color{white} \put(-125,80){\vector(1,0){15}} \color{white} \put(-135,77){B}$
\includegraphics[width=5.5cm]{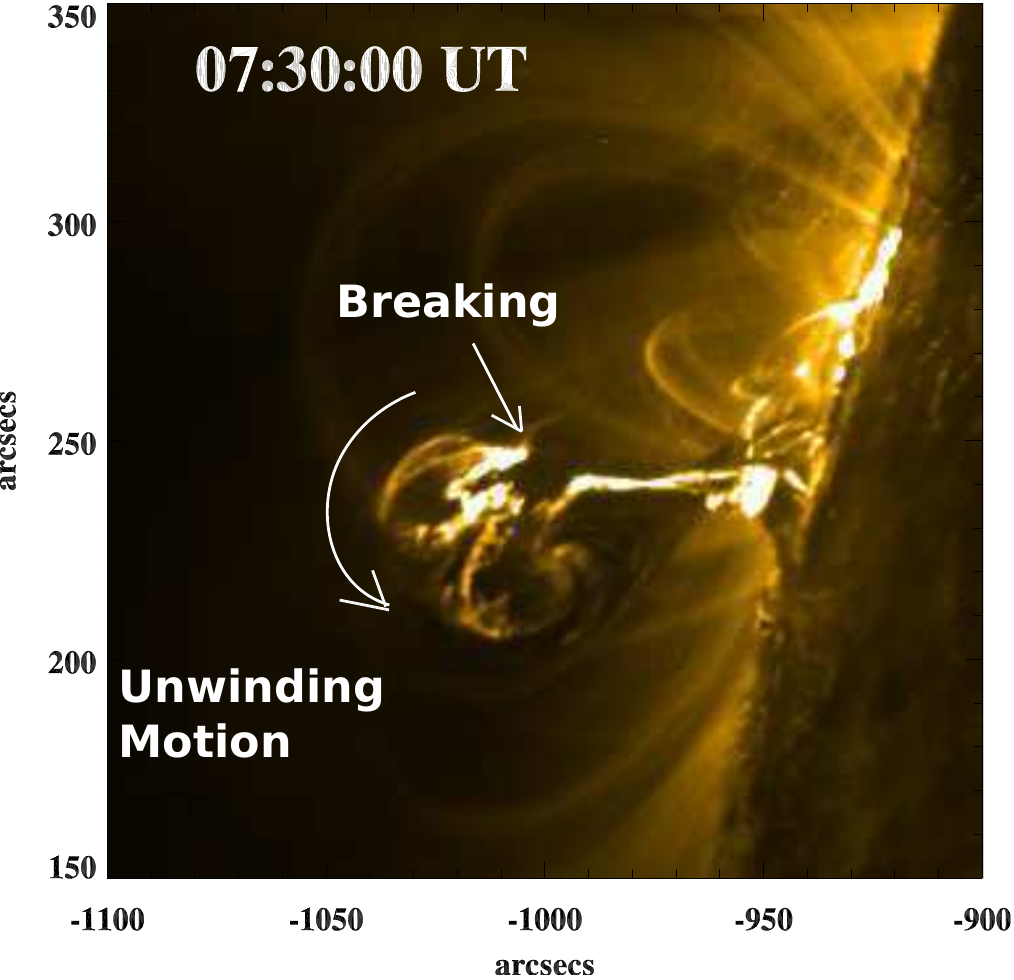}
}
\centering{
\includegraphics[width=5.5cm]{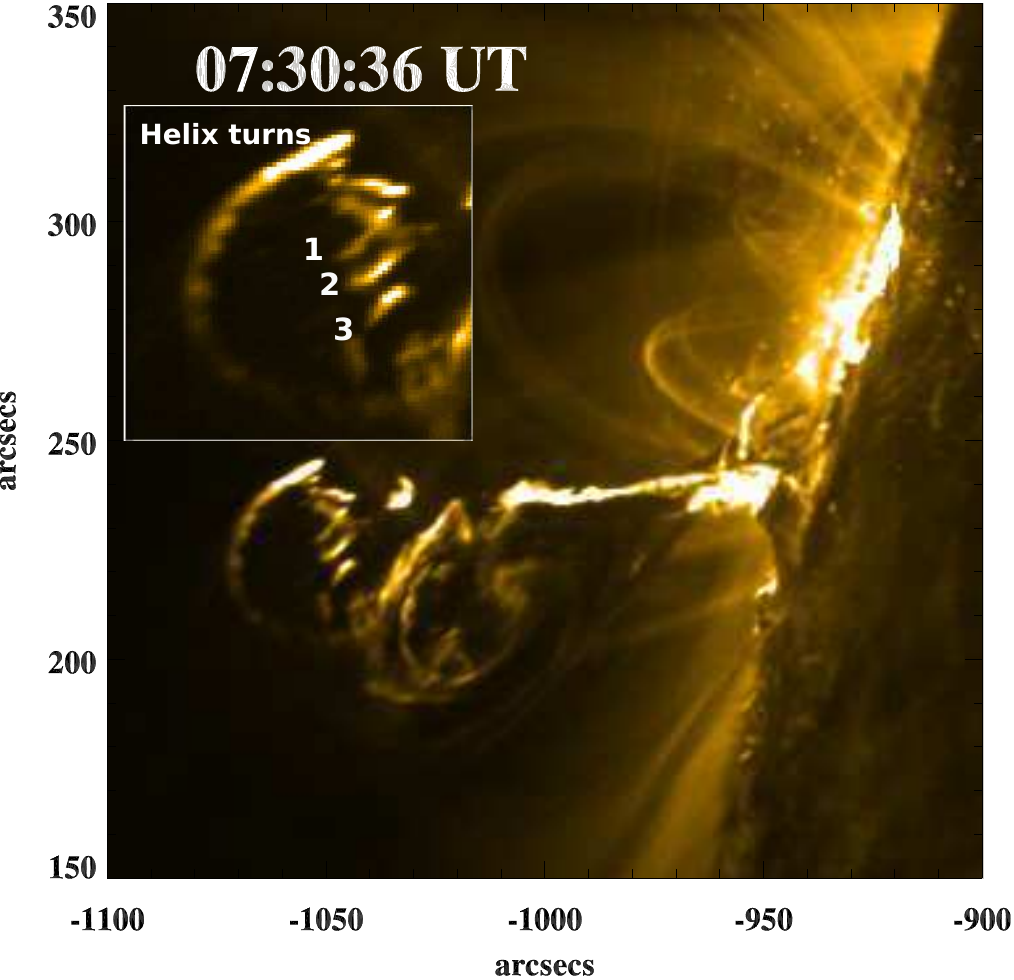}
\includegraphics[width=5.5cm]{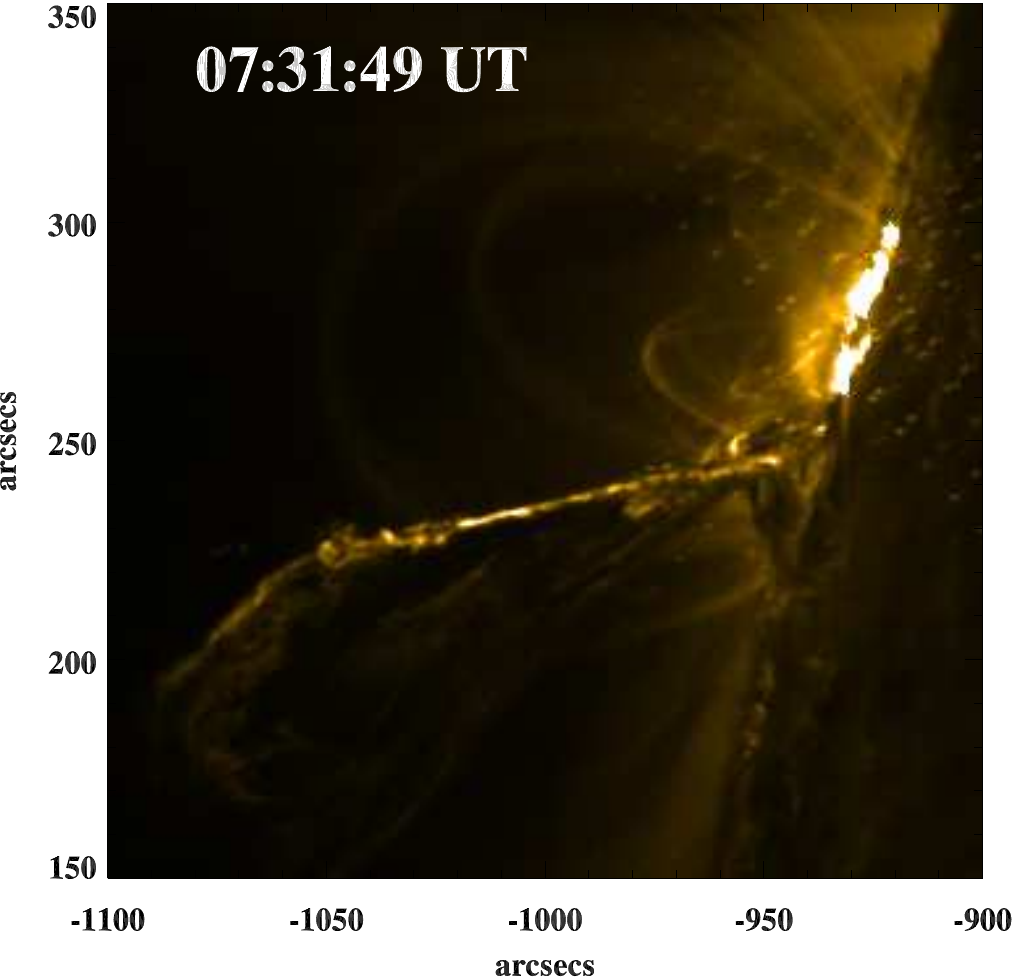}
}
\caption{SDO/AIA 171 (T$\sim$0.6 MK) \AA \ EUV images showing the development of kink instability, associated CME and flare. Top-right image shows the enlarged view of the helical brightening (indicated by arrow) in the flux rope structure. The enlarged view of three turns in the structure is shown in the bottom-left image.}
\label{aia171}
\end{figure*}

Figure \ref{aia171} displays the selected AIA 171 \AA \ EUV images of the corona. These images indicate the development of helical kink instability in the prominence which erupted at the onset of the CME. The occurrence of this instability implies that the magnetic field had the structure of a flux rope.  In interpreting the data of the event, we will make the assumption that the prominence was embedded in a  magnetic flux rope. Initially at 07:22:36 UT, we observe the dark prominence structure lying above the eastern limb in the center of coronal loop systems. Both legs of the prominence are visible (indicated by leg 1 and leg 2), extending along the north and south directions. The northern leg is longer/thinner than the southern one. It seems that most of the prominence mass is loaded on the southern leg. An  initial brightening is seen at the underside of the prominence below the apex at 07:24:12 UT. This shows the site of flare initiation. This brightening is the first signature of instability. It transiently displays a pattern indicative of a helical structure with $\sim$3-4 turns (shown by arrow in the enlarged part of the structure), which corresponds to a magnetic twist of $\sim$6$\pi$-8$\pi$. A loop-shaped structure developed at the apex of the prominence, following the brightening (heating) at the underside. This structure, denoted by blob `A', shows a rotational motion in the image plane in the counterclockwise direction, which is suggestive of an unwinding motion (refer to images at 07:27 and 07:29 UT and movie). In addition, another loop-like structure (indicated by `B') forms in the northern part of the prominence, also near the apex, from about 07:28 UT onward, and continues to grow with rising motion of the prominence.  Associated with the growth of blob `B', the northern leg of the prominence was detached at 07:30 UT. Blob `B' also exhibited rotational motion in the counterclockwise direction in the course of its fast rise after $\sim$07:29 UT. Additionally, the flare intensity started to rise with the rise-up of the prominence and maximizes at $\sim$07:35 UT. During the eruption, one could see a substructure in the center of blob `B' which was also indicative of a helix with $\sim$3 turns, indicated by 1, 2 and 3 in the enlarged part of the structure in the 07:30:36 UT frame. After the eruption of blob `B', blob `A' also expands with further rotational motion in the counterclockwise direction and stretches the overlying field lines (refer to AIA 171 \AA \ movie). The overlying loop structures above the prominence show an expanding motion during the eruption, which appears synchronous with the rise of the prominence. This suggests that these loops were located at or only slightly above the upper edge of a flux rope which supported the prominence. 

\begin{figure*}
\centering{
\includegraphics[width=6cm]{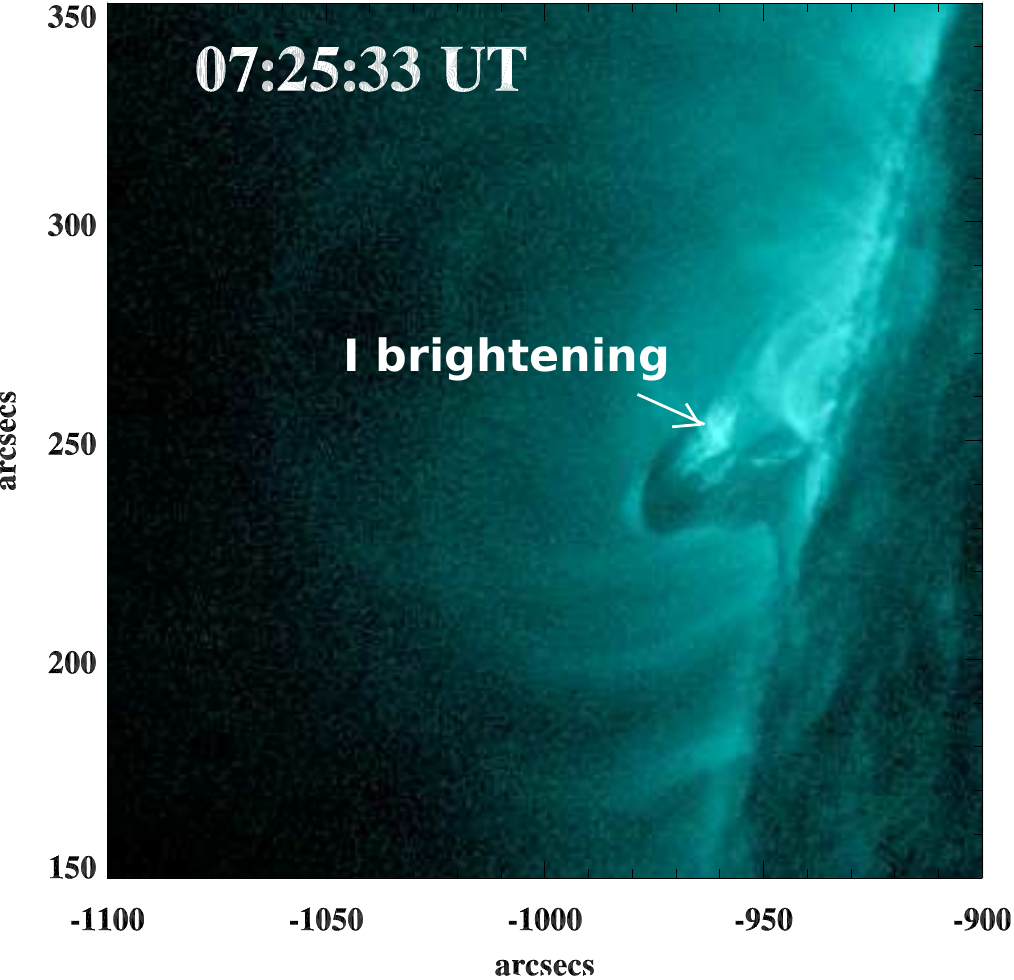}
\includegraphics[width=6cm]{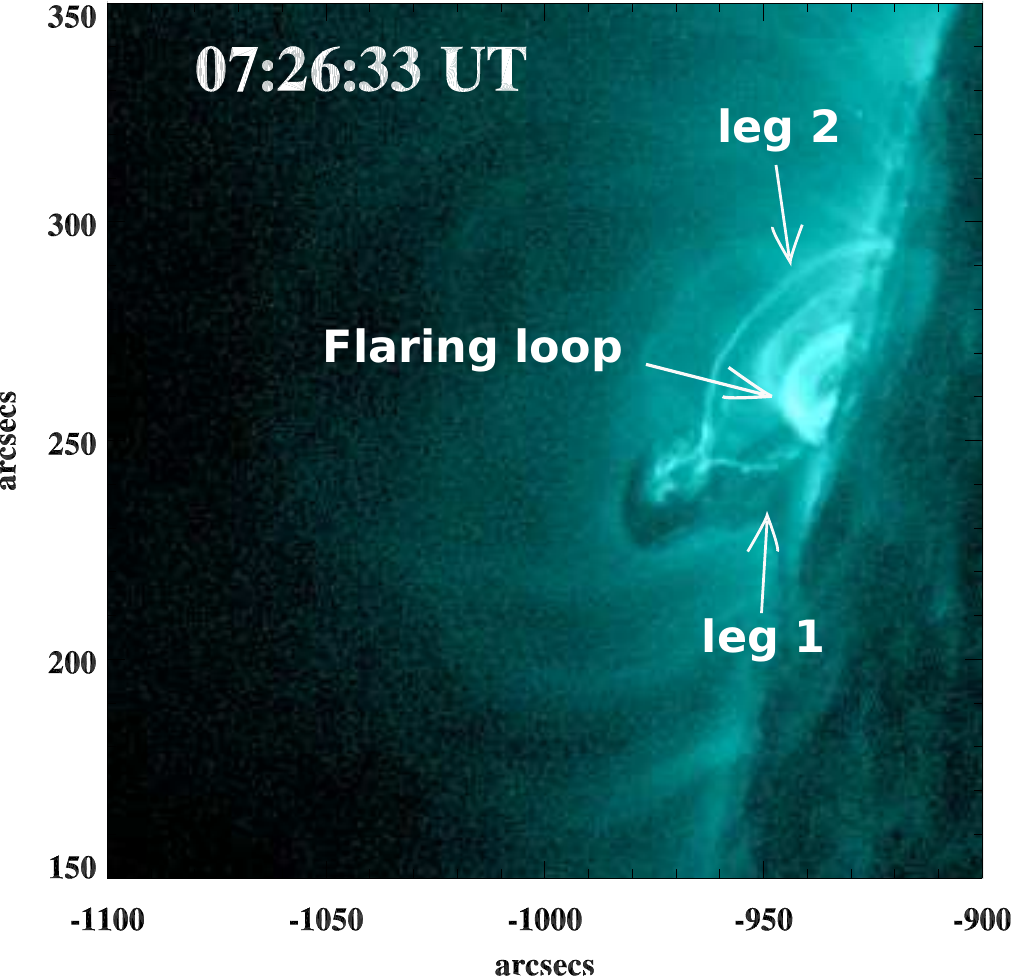}
}
{
\includegraphics[width=6cm]{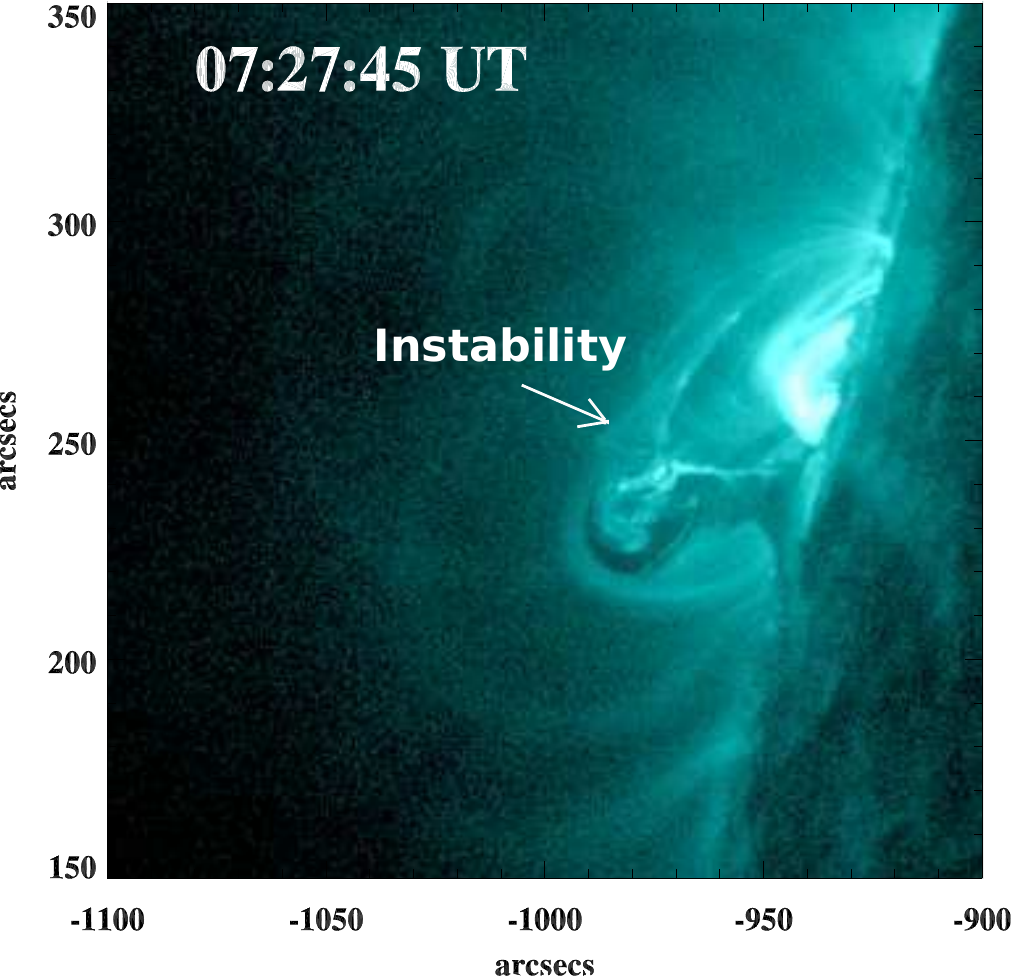}
\includegraphics[width=6cm]{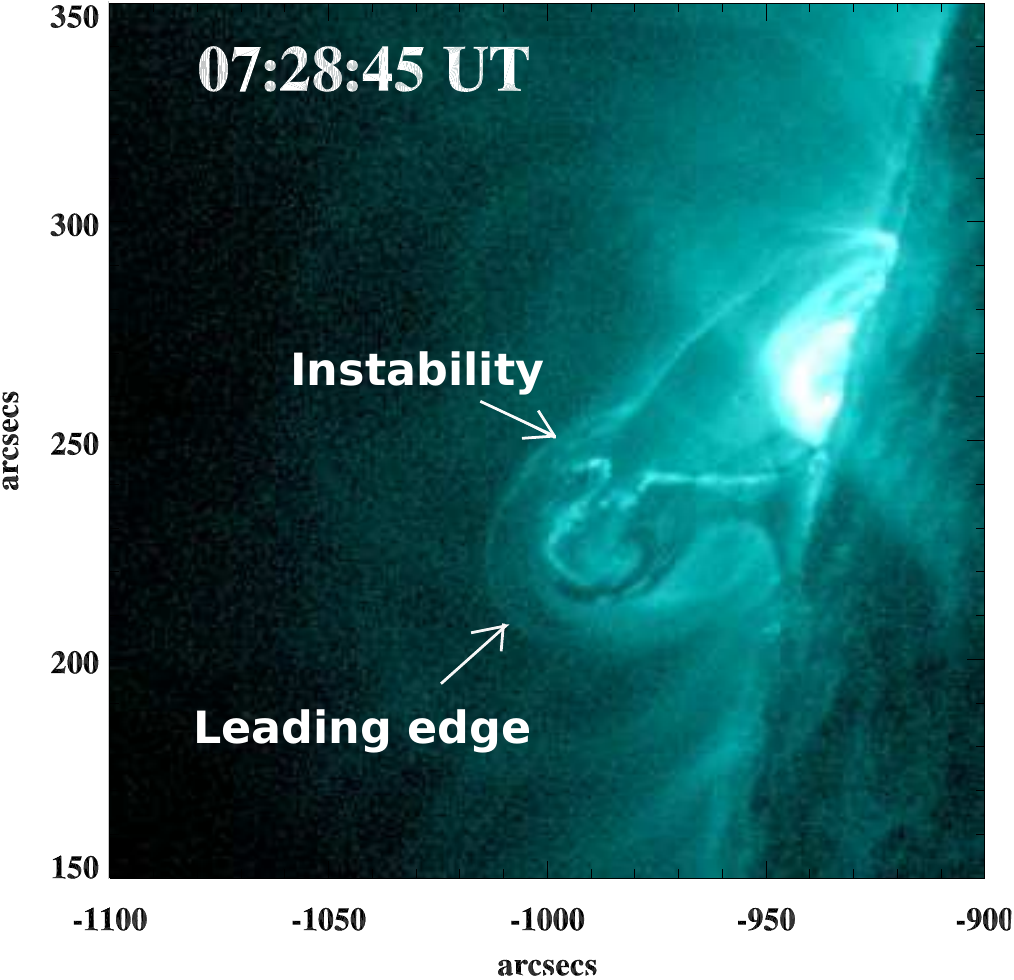}
}
{
\includegraphics[width=6cm]{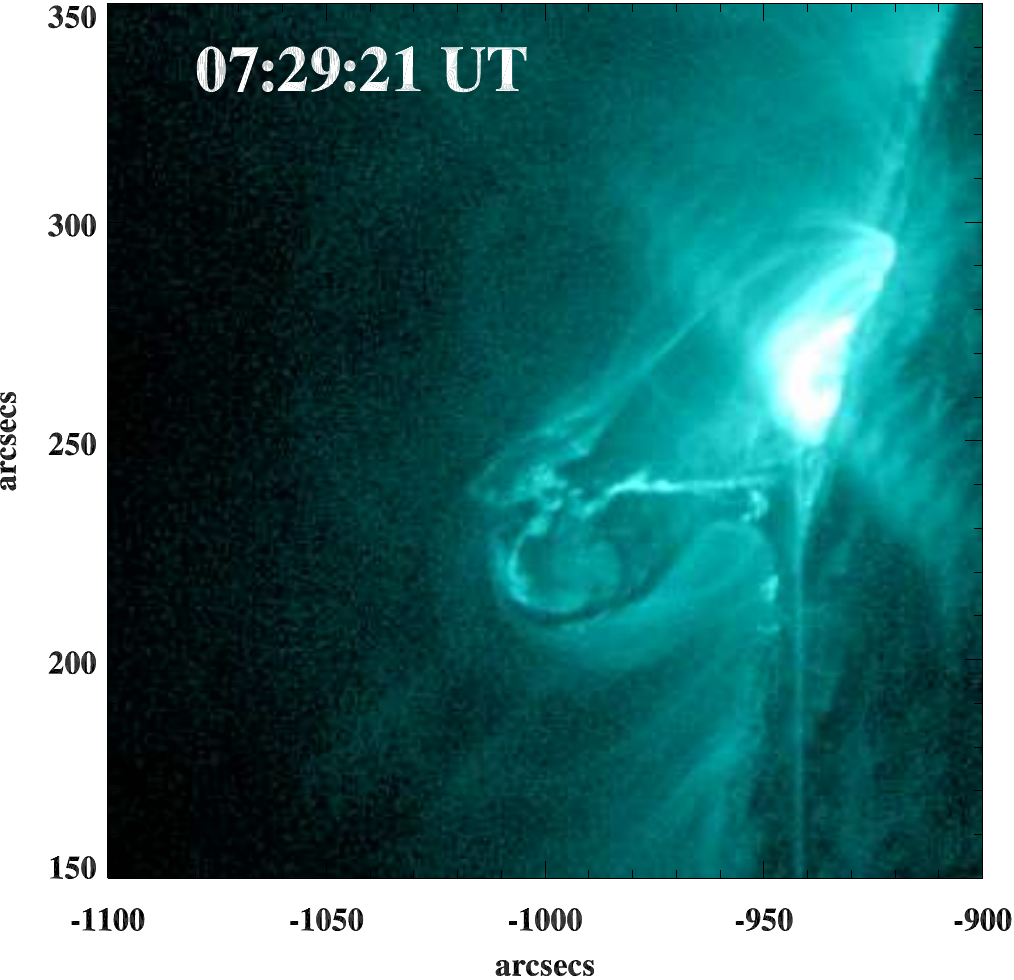}
\includegraphics[width=6cm]{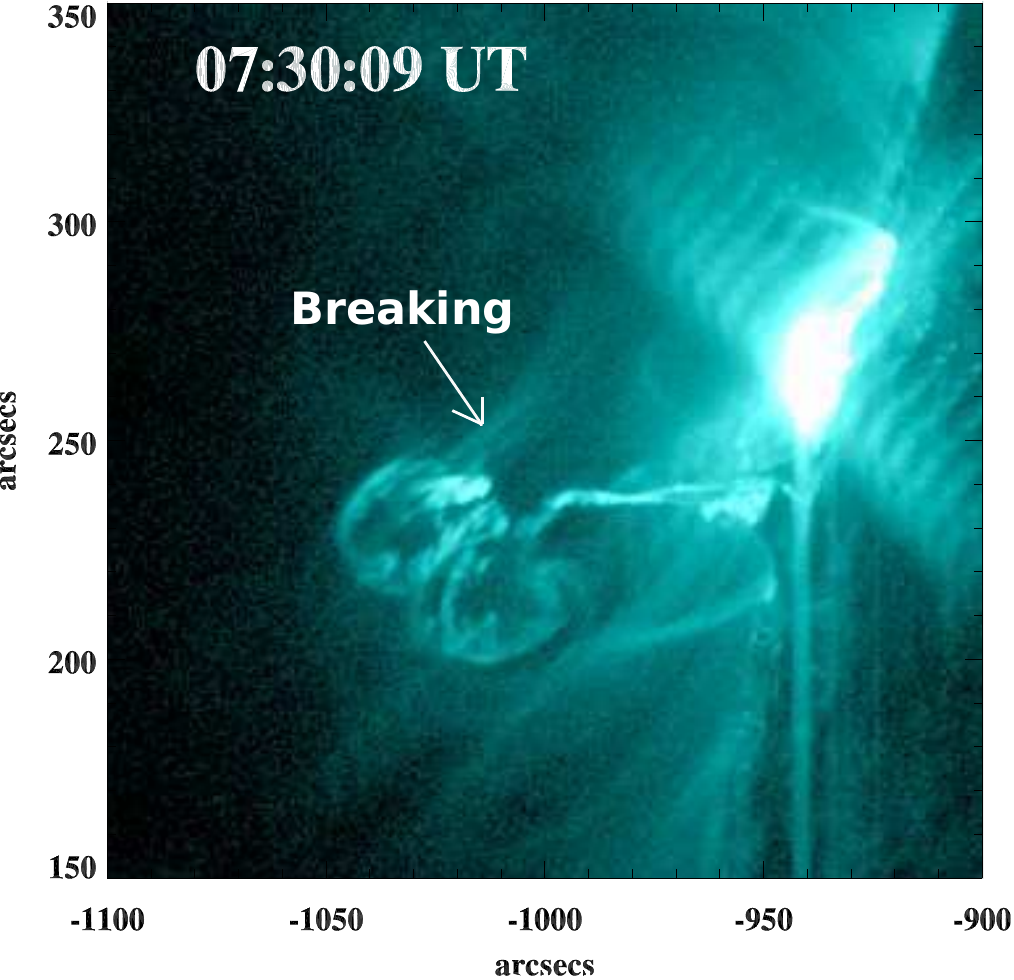}
}
\caption{SDO/AIA 131 (T$\sim$11 MK) \AA \ EUV images showing the development of instability, associated with CME and flare.}
\label{aia131}
\end{figure*}

In Figure \ref{aia131}, we have shown the selected images of AIA 131 \AA \ during the flare and CME initiation. These images correspond to a higher coronal temperature ($\sim$11 MK). The first image shows the initial brightening (indicated by arrow) at the underside of the prominence near its apex, as also seen in 171 \AA . The investigation of these images suggest the presence of a flare loop close to the footpoints of the prominence, beginning at 07:25 UT (shown by arrow in the 07:26:33 UT frame). It seems the formation of blobs is closely linked with the flare. The initial brightenings seen in the 171 \AA \ channel evolve into the formation of blob `A' and the first clear signs of the formation of blob `A' are simultaneous with the formation of first flare loop (both occurred during 07:25-07:26 UT). After the flare initiation, the fast rise of the soft X-ray and EUV flux commences simultaneously with the onset of the rapid ascent of blob `B' at $\sim$07:29:30 UT. 

\begin{figure*}
\centering{
\includegraphics[width=6cm]{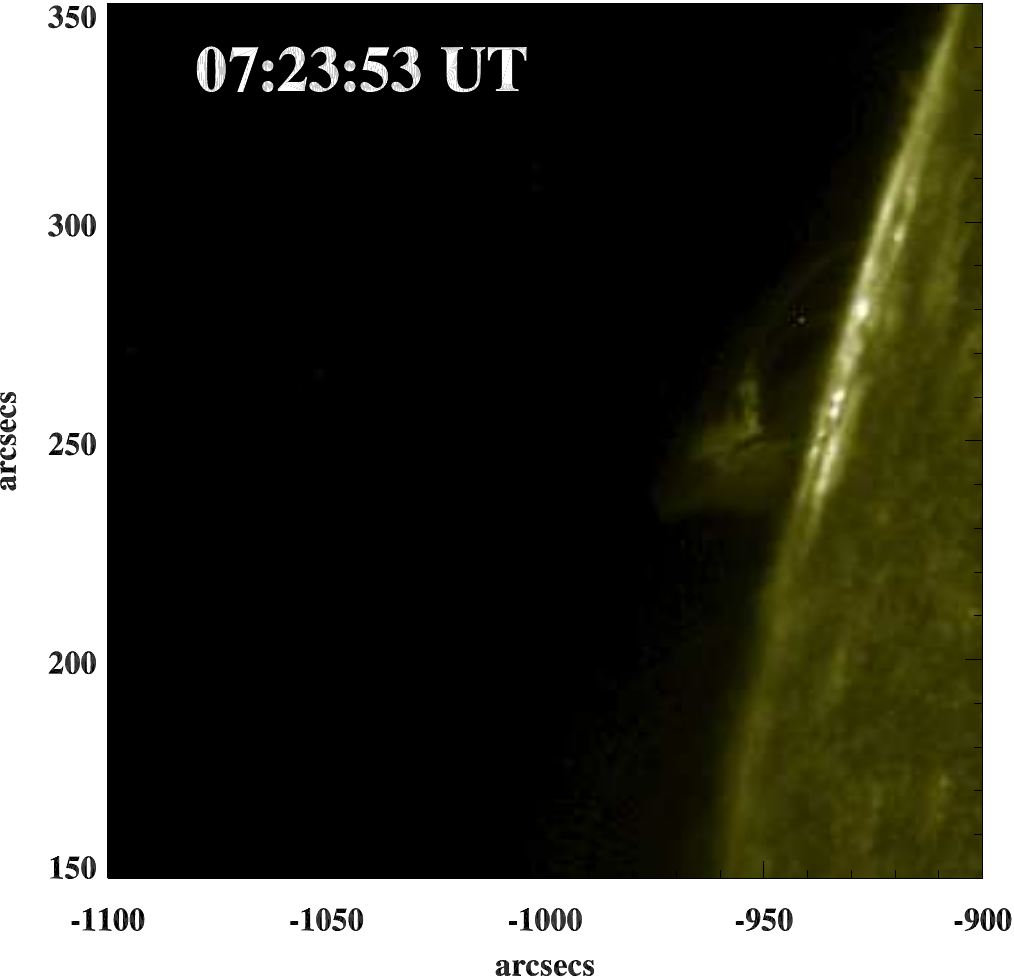}
\includegraphics[width=6cm]{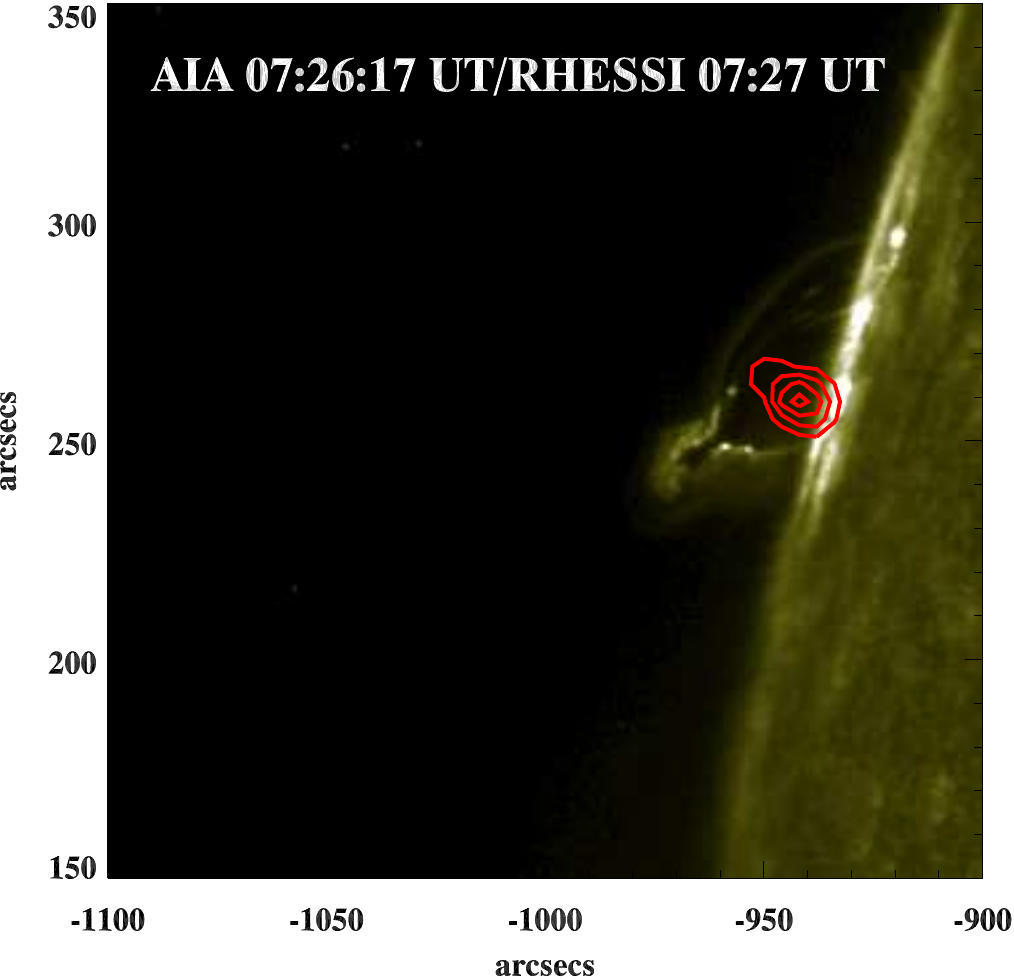}
}
{
\includegraphics[width=6cm]{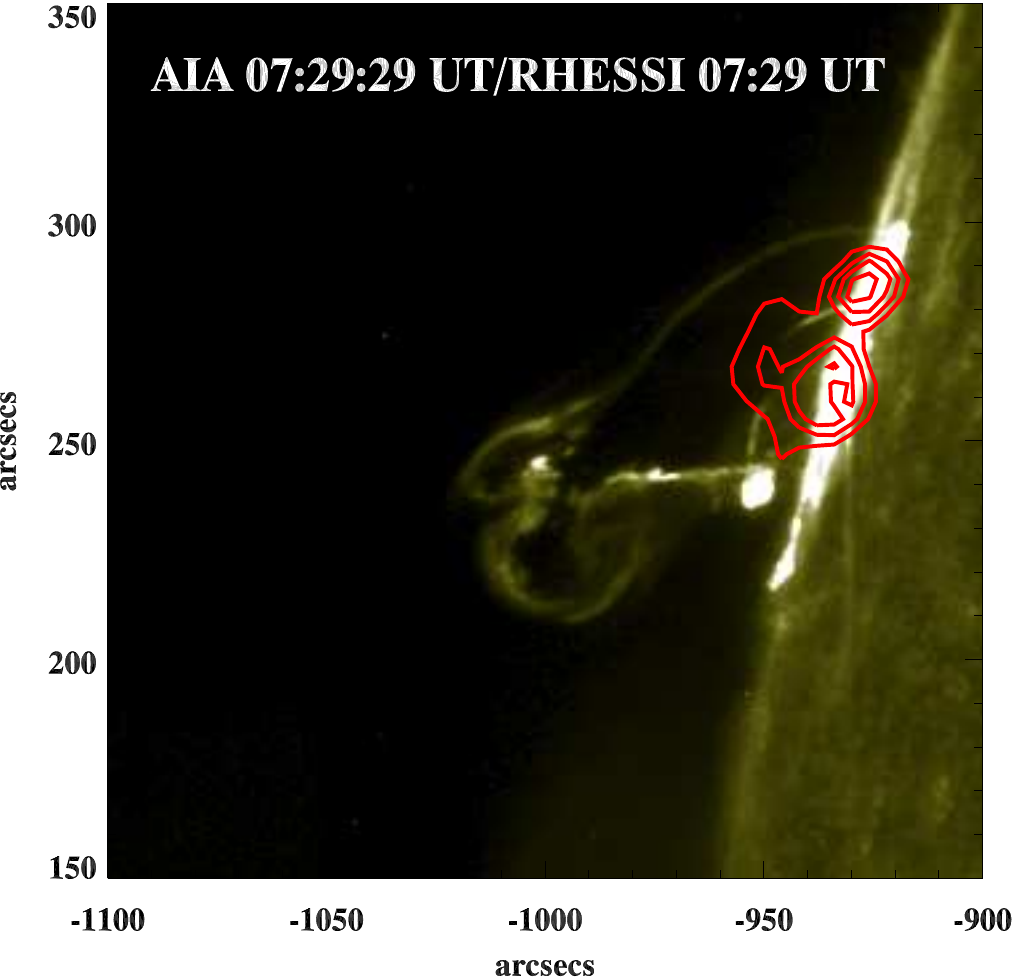}
\thicklines
$ \color{white} \put(-108,65){\vector(1,0){15}} \color{white} \put(-118,62){A}$
$ \color{white} \put(-115,84){\vector(1,0){15}} \color{white} \put(-125,80){B}$
\includegraphics[width=6cm]{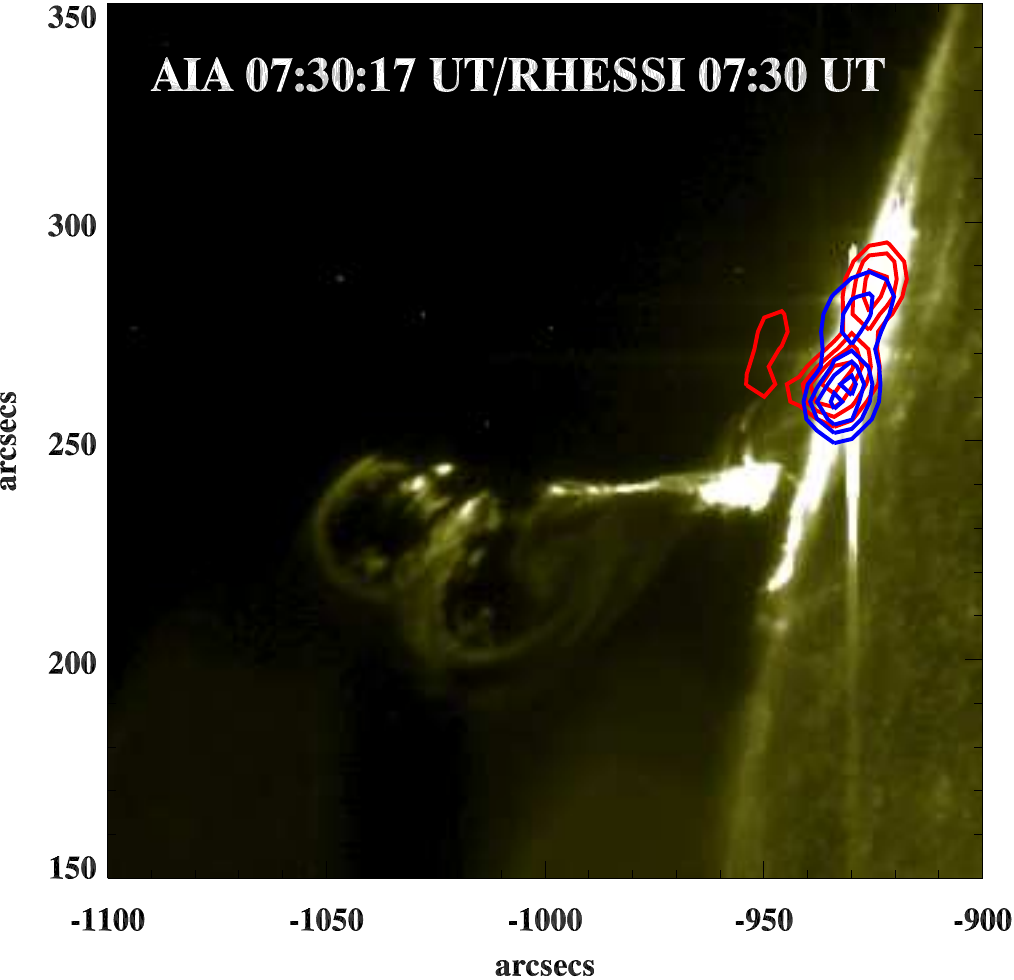}
}
{
\includegraphics[width=6cm]{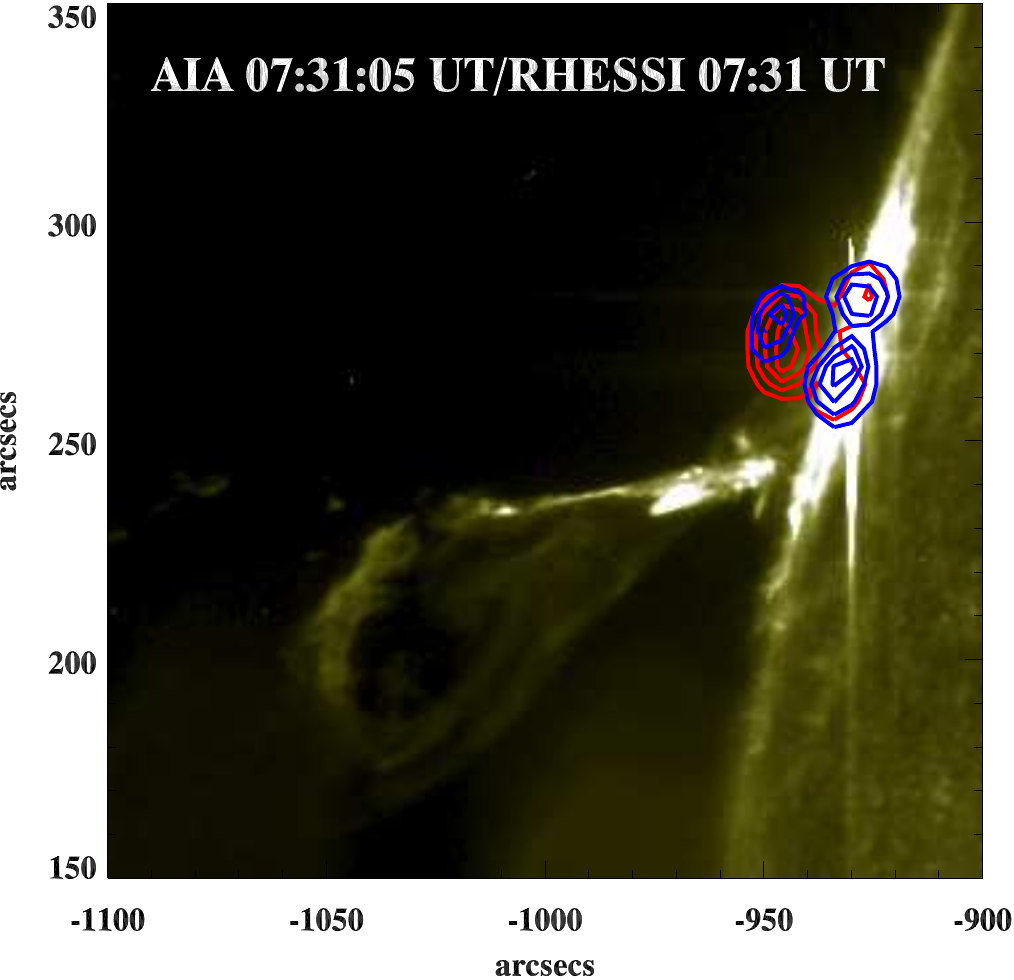}
\includegraphics[width=6cm]{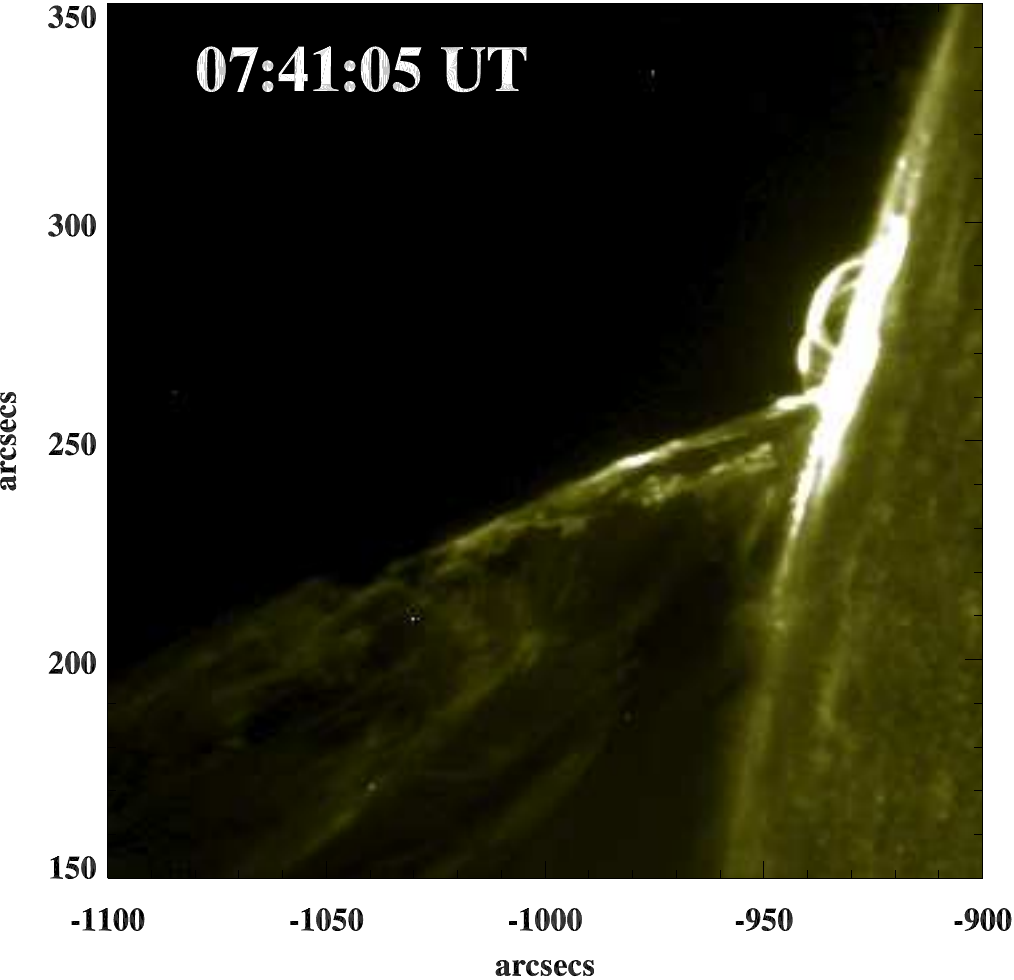}
\thicklines
$ \color{white} \put(-60,115){\vector(1,0){15}} \color{white} \put(-120,120){post-flare loops}$
}
\caption{SDO/AIA 1600 \AA \ (T$\sim$0.1 MK) images showing the development of kink instability, associated CME and flare. Some selected images have been overlaid by RHESSI hard X-ray contours. The contour levels are 30$\%$, 50$\%$, 70$\%$ and 90$\%$ respectively for both energy bands: 12-25 keV (red), 25-50 keV (blue).}
\label{aia1600}
\end{figure*}

We have used RHESSI \citep{lin2002} hard X-ray image data to investigate the particle acceleration. We reconstructed RHESSI hard X-ray images in 12-25 and 25-50 keV. We use `PIXON' algorithm for reconstruction of images. The image integration time in our case is 1 minute. Figure \ref{aia1600} shows the RHESSI hard X-ray contours overlaid on AIA 1600 \AA \ images. Red contours represent the hard X-ray sources in 12-25 keV whereas blue one indicate 25-50 keV energy bands. At 07:26:17 UT, we see single hard X-ray source in 12-25 keV near the southern foot of the prominence during the flare initiation. During the development/detachment of blob `B', the footpoint sources of flare are seen in both energy bands during 07:29 to 07:31 UT. This suggests that the acceleration of nonthermal electrons is closely related with the eruption of blob `B'. Later on, during the ejection of blob `A', we could see another coronal source at  07:30-07:32 UT, which may be coronal loop-top source. Post-flare loops can be seen in the image at 07:41:05 UT, which connect both footpoints of the flare as seen in RHESSI images. Therefore, the RHESSI images confirm to the standard picture of hard X-ray (HXR) sources formation at the feet of the flare loops in the higher energy range and HXR source formation all along the flare loop in the lower energy range. 

\begin{figure*}
\centerline{
\includegraphics[width=14cm]{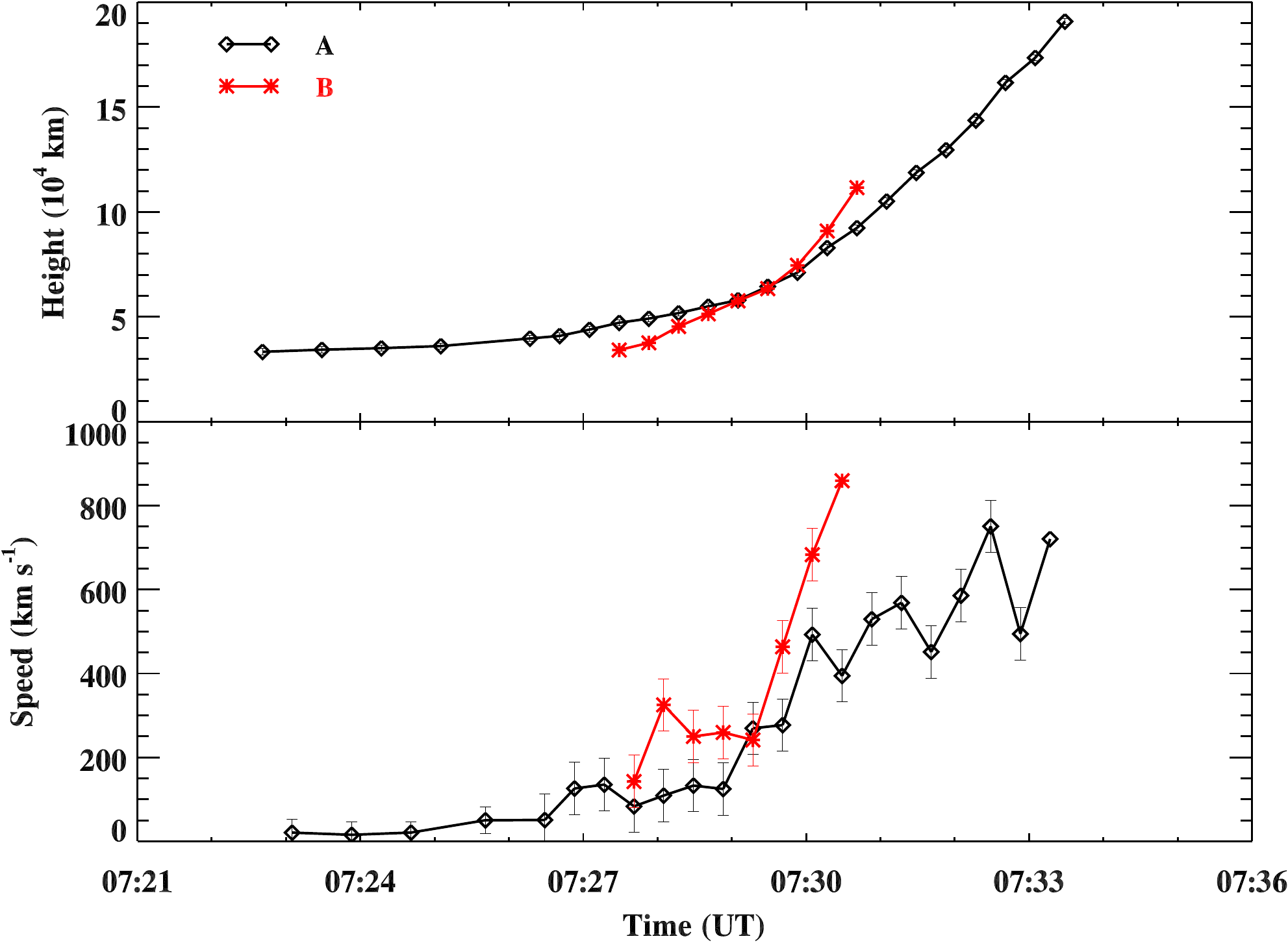}
}
\caption{Height-time and speed profiles of the leading edge of blobs `A' and `B' derived from AIA 1600 \AA \ images.}
\label{ht}
\end{figure*}

\begin{figure*}
\centerline{
\includegraphics[width=12cm]{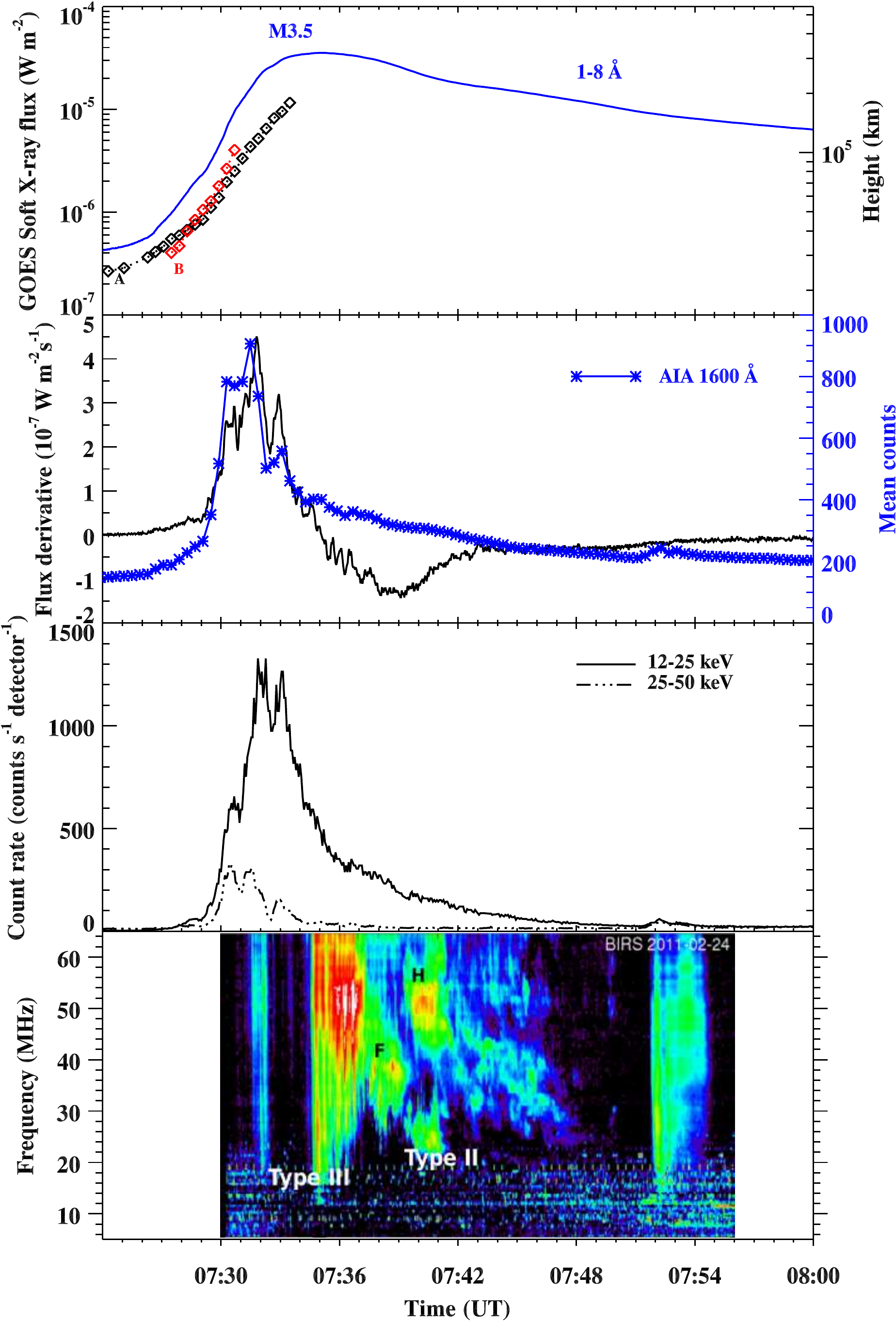}
}
\caption{GOES Soft X-ray flux profile along with blobs (`A' and `B') height-time plots, soft X-ray flux derivative along with AIA 1600 \AA \ flux, RHESSI hard X-ray flux profiles (12-25 and 25-50 keV) and dynamic radio spectrum (5-65 MHz) observed at BIRS (Bruny Island, Tasmania) showing the type III and type II radio bursts on 24 February 2011.}
\label{flux}
\end{figure*}

In order to study the kinematics of the rising prominence, we perform the distance-time measurements of the leading edge of both blob structures using AIA 1600 \AA \ images. We took the reference point at the base of the structures. We assume the measurement error of 2$\arcsec$ in the calculation. Blob `B' started to evolve at $\sim$07:28 UT, was strongly accelerated at $\sim$07:29:30 UT, and left the field of view after $\sim$07:31 UT. We could track blob `A' from the beginning to the end of the flare. Figure \ref{ht} displays the height-time profile of these two blobs `A' and `B' in the top panel. The corresponding speeds of both structure have been plotted in the bottom panel. The structure `A' shows the activation phase of the eruption before 07:29 UT, given by slow rise with an average velocity of $\sim$100 km s$^{-1}$, and subsequently it shows a strong acceleration. The speed of the structure during this time rises to $\sim$800 km s$^{-1}$. The blob structure `B' initiated with the speed of $\sim$200-300 km s$^{-1}$ till 07:29 UT and later it is also accelerated strongly to reach a speed of at least $\sim$900 km s$^{-1}$. Therefore, we could investigate the kinematics of the both blobs in the structures due to high resolution AIA images, which shows the activation and the eruption phase of the CME. The speed profile of blob `A' shows the quasi-periodic oscillations after 07:27 UT. The estimated period of these oscillation was $\sim$2 minutes.

In order to investigate the close relation between the eruption and associated particle acceleration, we have plotted different flux profiles (soft X-ray, EUV and radio) against the height-time profiles of the blobs `A' and `B' (see Figure \ref{flux}). The top panel shows the GOES soft X-ray profile in 1-8 \AA \ wavelength band and height time profile of both structure. The first signs of the helical kink instability (i.e. the formation of blob `A') were observed between 07:25-07:26 UT. This was before the fast rise of the GOES light curve, which began near 07:29 UT. The steepest rise of the soft X-ray flux began nearly simultaneously with the onset of the main acceleration of blobs `A' and `B'. This shows the close association between these two phenomena. The second panel depicts the soft X-ray flux derivative and average counts of AIA 1600 \AA \ EUV images (shown in Figure \ref{aia1600}). It is interesting to note that both are showing three peaks with one to one correlation associated with the acceleration of nonthermal particles towards the footpoints of the flare. This EUV emission mostly related the flare ribbons/footpoints.  The third panel displays the hard X-ray flux profiles in 12-25 and 25-50 keV. After the careful comparison, we found three peaks of hard X-ray flux coincide with the soft X-ray derivative and EUV flux peak in the second panel. This implies that the accelerated electrons that produce the hard X-ray also heat the plasma that produce the soft X-ray, obeying the Neupert
effect \citep{neupert1968}.  The bottom panel shows the dynamic radio spectrum (6-65 MHz) observed at BIRS (Bruny Island, Tasmania).  We observe the type III radio burst during 07:34-07:37 UT at flare maximum, which indicates the existence of open field lines in the flare volume. Type II radio burst was also observed during 07:37-08:05 UT, which is the signature of coronal shock. This shock is most likely associated with the piston driven shock by CME. The estimated speed using drift rate of the type II was $\sim$1283 km s$^{-1}$. Therefore, the event showed significant amount of particle acceleration during the eruption process.

\begin{figure*}
\centerline{
\includegraphics[width=7cm]{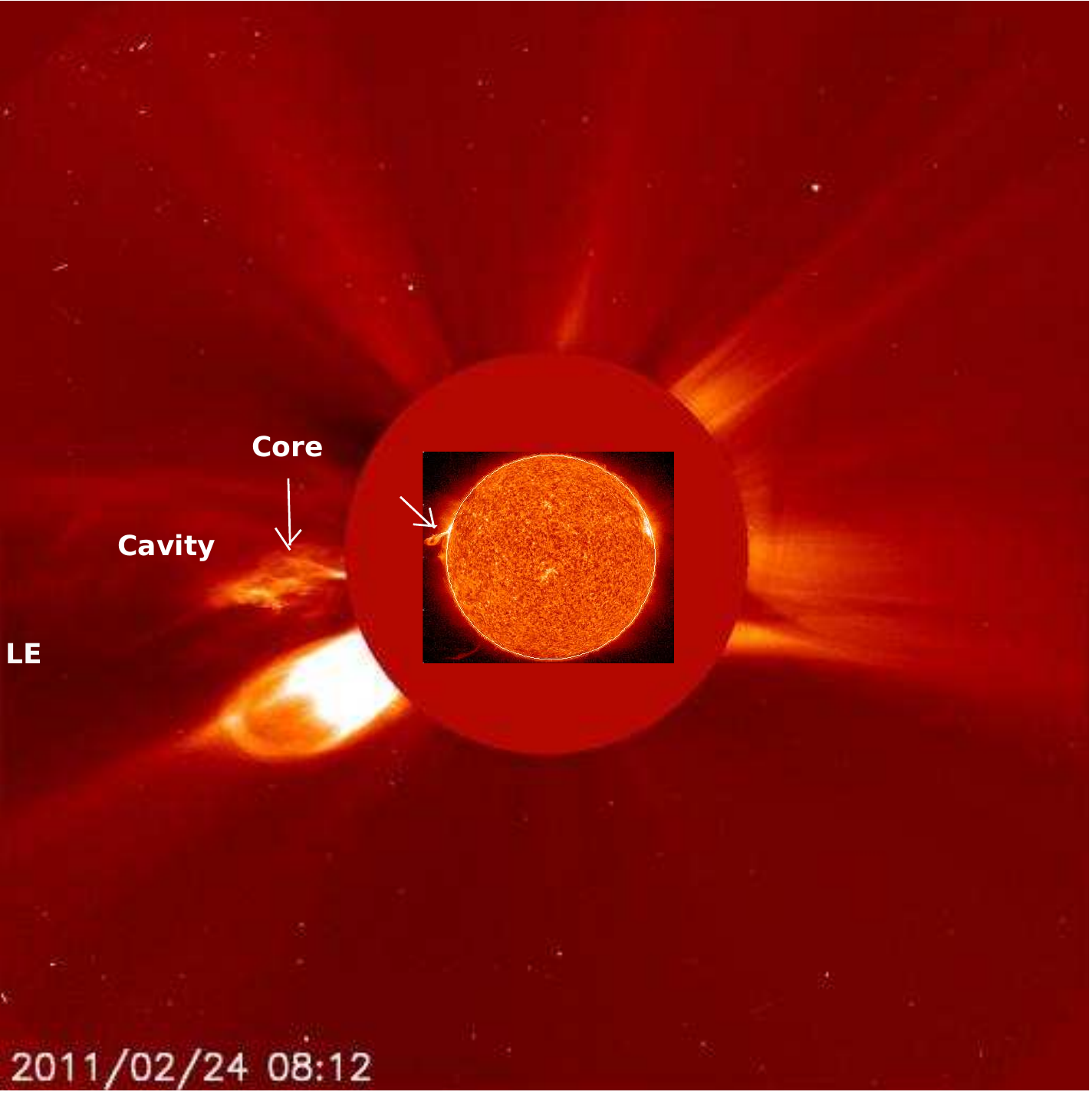}
\includegraphics[width=7.2cm]{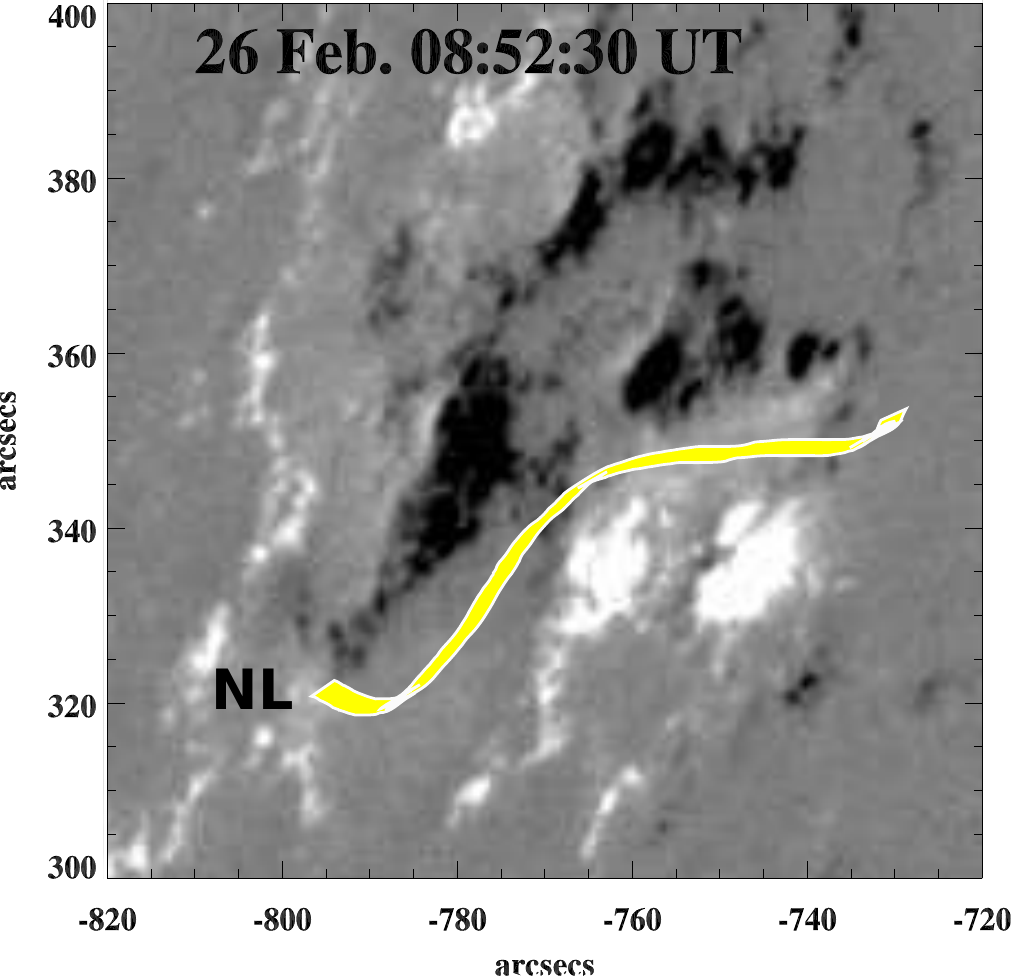}
}
\caption{Left: SDO/AIA 304 \AA \ (T$\sim$0.05 MK) image (center) showing the erupted flux rope at 07:33:08 UT and SOHO/LASCO white light image of the associated CME (indicated by an arrow). Right: HMI magnetogram showing the morphology of active region NOAA 11163 on 26 February, 2011. It shows `S' shaped neutral line along which flux rope structure was activated (indicated by yellow color). }
\label{cme}
\end{figure*}

LASCO C2 and C3 \citep{bru1995} observed the CME associated with this eruption. The left panel of Figure \ref{cme} displays the composite image of AIA 304 \AA \ combined with LASCO C2 white light image. The AIA 304 \AA \ image shows the erupting prominence at the eastern limb indicated by an arrow and the LASCO C2 image displays the CME in the same direction. This CME is a good example of 3-part CME structure, which shows the leading edge, cavity and core (shown by arrow). The leading edge is the overlying coronal loop which was expanding during the flux rope eruption in AIA 171 \AA \ images (refer to Figure \ref{aia171}). The structure of the prominence in the core of the CME and the existence of a CME cavity support the interpretation of the erupting structure as a flux rope. 

\begin{figure*}
\centering{
\includegraphics[width=10cm]{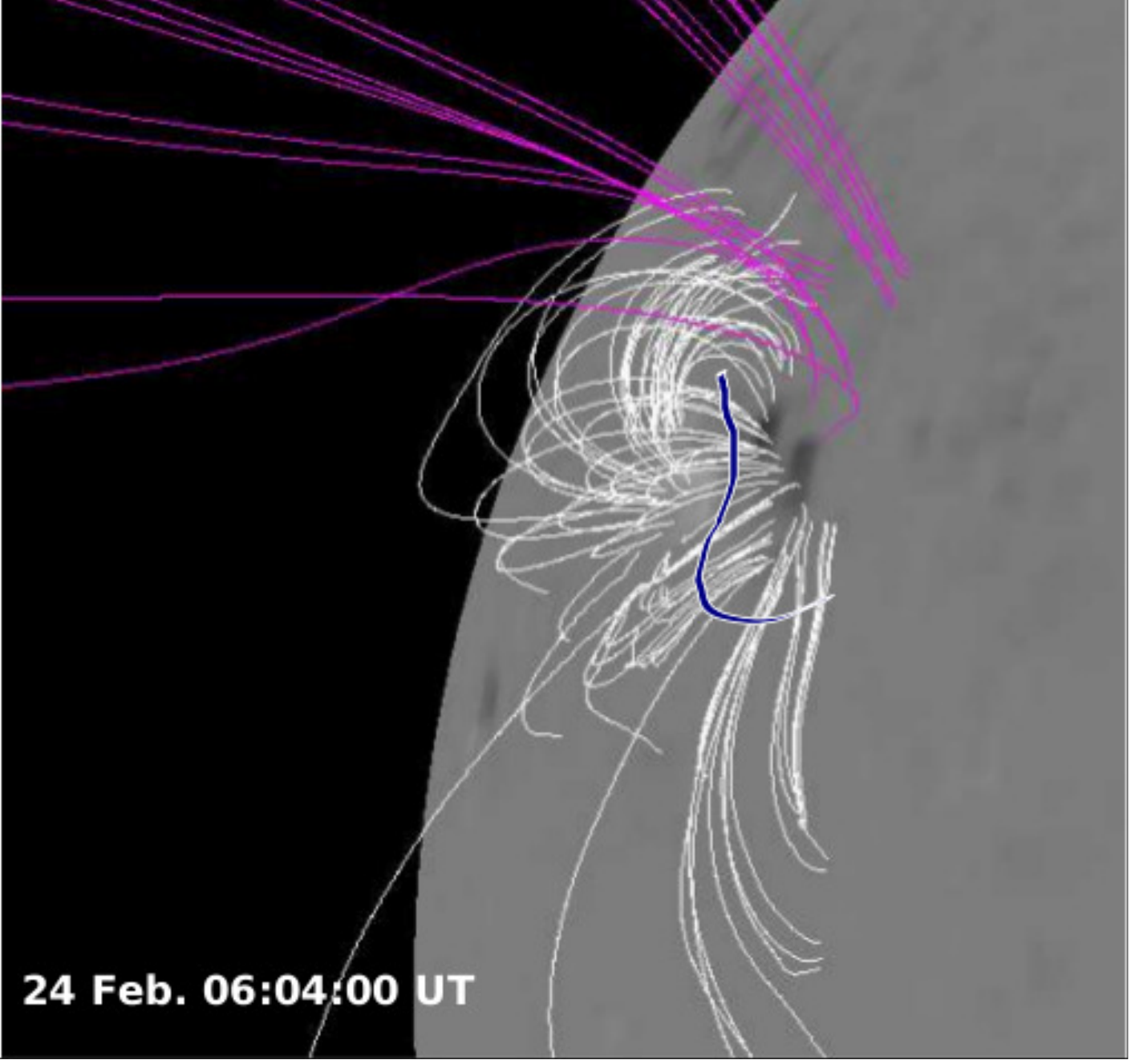}
}
\caption{PFSS extrapolation of the active region NOAA 11163 at 06:04 UT on 24 February 2011. The magnetogram has been rotated towards west to view the field lines. White lines show the closed field lines whereas red one indicate the open field lines in the active region. The neutral line is indicated by blue line, which is the site of CME initiation.}
\label{pfss}
\end{figure*}

We use the SDO/HMI magnetogram to view the magnetic morphology of the active region. The right panel of Figure \ref{cme} shows the HMI magnetogram observed two days after the CME, on 26 February 2011. We can see the `S' shaped polarity inversion line and this geometry is favorable for flux rope eruption \citep{cho2009}. The northern footpoint of the prominence was anchored in the negative polarity whereas the southern footpoint in the positive polarity field region. For investigating the overlying magnetic field environment of this active
region, we have used the potential field source surface (PFSS) extrapolation
\citep{alt1969,sch1969} before the flare event at 06:04 UT on 24 February 2011 (see Figure \ref{pfss}). The PFSS model uses full-sun surface-flux map  by assimilating MDI or HMI magnetograms \citep{sch2003}. We have rotated the magnetogram towards west to view the  extrapolated field lines in the active region. The magnetic field extrapolation is in agreement with the field lines shown in AIA EUV 171 \AA \ image (top-left image in Figure \ref{aia171}). The neutral line has been indicated by blue color, which matches well with the location of the CME initiation.   

\section{INTERPRETATION}

\subsection{Occurrence of the Helical Kink Instability}
In the present event, we observed the following evidences of the helical kink instability in the flux rope.
 
(i) The strong indication in favor of helical kink instability is the formation of blobs `A' and `B' in the course of the eruption. After the blobs `A' and `B' were fully developed, they jointly appear to form a helix with nearly two turns. This is indicated in the AIA images from about 07:28:30 UT onwards and most clearly expressed during $\sim$07:29:30--07:30:30 UT. This apparently helical structure comprises the main body of the prominence. Therefore, it is quite likely that it comprises the magnetic axis of the assumed flux rope. In other words, these images suggest that the magnetic axis of the flux rope has taken a helical shape, which is the key property of the helical kink instability. Furthermore, since the STEREO-B images show that the northern footpoint of the prominence was located closer to the Earth than the southern footpoint, each of the loops at the edge of blobs `A' and `B' must have the shape of a left-handed helical curve. The combined loops of `A' and `B' thus formed a section of a left-handed helix. This chirality is consistent with the well known hemispheric rule and supports the interpretation in terms of helical kink instability. The counterclockwise rotating motion of these blobs in the image plane is consistent with the unwinding motion of a large-scale helix. Therefore, the structure (A+B) in itself is a strong evidence for the occurrence of helical kink instability in the flux rope.

(ii) The images additionally contain two substructures which may indicate strongly helical field lines in the erupting structure.  The observations yield indications of twisted helical substructures showing $\sim$3-4 turns, which is equal to the magnetic twist of 6$\pi$-8$\pi$, well above the threshold value of the kink instability \citep{fan2003,fan2004,kliem2004,torok2003,torok2004}. The first bright helical structure was indicated below the prominence in AIA 171 \AA \ images (07:24:12 UT) prior to the eruption. The second structure was visible in the middle of blob `B' after 07:30:12 UT  i.e., after the main energy release in the event had started. On the other hand, we interpret these substructures as a weak indications of the helical kink instability due to the following reasons: (a) both structures do not occur at the right place i.e. at the magnetic axis of the assumed flux rope. Prominences are thought to be material supported against gravity in dipped sections of the field lines. In a flux rope, dipped sections exist between the bottom of the rope and the magnetic axis somewhere near the geometric middle of the rope. This means that the magnetic axis always runs very near the upper edge of the visible prominence material, or even above the upper edge. At the location of first brightening the prominence is rather thick. Therefore, this brightening must lie considerably below the magnetic axis of the rope. If the bright streaks would show us the full cross-section of the helix (as suggested by the schematic in Figure 1, top right), then the complete helix would lie below the magnetic axis of the assumed flux rope. Then, this would be a small separate flux rope. It appears unlikely that such a tiny substructure, located at a distance to the magnetic axis of the rope, can force the whole flux rope to develop a helical shape and erupt. Alternatively, there may be another possibility. The bright streaks illuminate only a short section of 4-5 field lines that actually wind around the flux rope axis, but with a considerable distance to the axis (i.e., they have their highest point considerably above the visible prominence material). This is possible because the bright streaks in 171 \AA \ image at 7:24:12 UT are not connected to form a full helix. It could be possible that the brightening shows that the external flux is added to the flux rope by reconnection under the rope (similar to tether-cutting reconnection) and that the new flux is wound about the rope, in this way increasing the total average twist. For this interpretation, we have to assume that winding field lines (twist) were added to a preexisting rope by reconnection under the rope in order to interpret this brightening as a support for the helical kink instability. (b) The first structure is clearly visible only very transiently, in only 1--2 frames and the second structure becomes visible only after the main energy release. Therefore, in our opinion, these are the weaker indications for the occurrence of the helical kink instability than the helix indicated by the combined blobs `A' and `B'. 

\subsection{Estimations for the Ballooning Instability}
 We have only single observation of ballooning instability reported by \citet{shibasaki1999} in flaring loop using radio observations. After that no observation was reported related to the ballooning instability. In this section, we test the 
possibility of the ballooning instability in the prominence for the generation of semicircular plasma blobs.
The condition of the ballooning instability 
for the top of symmetric loop derived by \citet{shibasaki2001} is
\begin{equation}
    \beta_{T} > \frac{2r}{R} \;
\end{equation}

Where `r' and `R' are the loop radius and curvature radius of the loop respectively. 
Ballooning instability is expected in a symmetric loop, if the thermal beta value 
exceeds 2 times of the inverse aspect ratio of the loop. 

Using AIA observations, we tried to estimate whether the condition of the ballooning 
instability is fulfilled in the present event. If we take plasma density at the prominence apex
 n$_{e}$$\sim$5$\times$10$^{10}$ cm$^{-3}$ \citep{asc2004}, and assuming the fully ionized plasma below the apex of the prominence 
(where brightening starts in AIA 171 and 131 \AA \ images). The gas pressure is given by 
\begin{equation}
P=2n_{e}k_{B}T \;
\end{equation}

Where, k$_{B}$ is Boltzmann constant. However, the response temperatures for AIA 171 and 131 \AA \ respectively are 
0.6 and 11 MK. We used GOES soft X-ray flux profiles in 0.5-4 and 1-8 \AA \ wavelength bands to estimate the plasma 
temperature during the formation of plasma blobs at the apex of the prominence \citep{garcia1994}. The plasma was heated upto $\sim$5-6 
MK during 07:24-07:26 UT. Therefore, we use T=5-6 MK in the above expression. The plasma $\beta_{T}$ is given by
 
 \begin{equation}
 \beta_{T}=\frac{P}{B^2/8\pi}=\frac{6.92\times10^{-15} n_{e}(cm^{-3}) T(K)}{B^2(gauss)}
 \end{equation}

We assume B$\sim$50 G at the apex of the prominence \citep{zirin1961,sakai1982}. 
Thus the estimated value of plasma $\beta_{T}$ is $\sim$0.691 to 0.830.

The distance between the footpoints of the prominence is $\sim$60$\arcsec$ 
(from AIA 131 \AA \ image at 07:26:33 UT). Therefore, the typical radius of 
curvature (R) of the prominence (assuming semicircular shape) is $\sim$30$\arcsec$.
 However, the southern leg (leg 1 in Figure \ref{aia131}) was more thicker than the northern leg (i.e. leg 2) of the prominence. But, the 
 instability was mainly occurred in the northern leg (formation of plasma blob `B'), therefore we used the 
  approximate thickness of the northern leg in our estimation. The thickness of northern leg of the prominence is $\sim$6$\arcsec$ (refer to AIA 
 131 \AA \ image at 07:26:33 UT). This gives the approximate value of radius of the northern leg (r) 
 $\sim$3$\arcsec$. Using above measurements, the estimated value of 2r/R is $\sim$0.20.
 Therefore, from the above calculations the instability condition according to 
 equation (1) plasma $\beta_{T}$$>$0.20 is satisfied.  Based on the above assumptions, this event is in agreement with 
 the ballooning instability criteria, which may be generated at the prominence apex.
However, it is very likely that the flux rope which hold the visible prominence material 
have cross-sections that are considerably larger than the prominence (e.g., \citealt{bobra2008},
\citealt{gibson2010}).  In addition, 
 a set of overlying loops visible in AIA 171 \AA \ images move synchronous with the prominence 
  from the very beginning of the eruption (07:21 UT onwards). This motion is similar to the motion 
  of overlying loops in the events described by \citet{sch2008}, who demonstrated the tight association with the motion of the erupting prominence. This 
  motion represents an indication that the unstable flux rope fills much, or possibly even all, of the space 
  between the prominence and those overlying loops. Therefore, the flux rope may have been considerably thicker.
 In this case the ballooning instability does not occur. Since the estimates for the radius of the flux rope fall in a very large range, a definite statement about the occurrence of the ballooning instability is not possible at this point. 

 The growth time of the ballooning instability is also dependent upon the radius of the flux rope. According to \citet{shibasaki2001}, the linear growth time of the ballooning instability, which is
\begin{equation}
 \tau(s)\sim100\surd{lR/T}
 \end{equation}
Where, `R' is a curvature radius of the loop measured in the unit of 10 Mm and `l' is 
a density gradient scale length (roughly equal to loop thickness) perpendicular to 
the field in the unit of 10 Mm. T is the plasma temperature in MK. 
Using AIA observations (171 and 131 \AA), $R\sim30\arcsec$,  
thickness of the northern leg $\sim$6$\arcsec$ and T$\sim$~5-6 MK (from GOES measurements). 
Using these parameters in the above equation, the linear growth time for the 
above observations is $\sim$40 s.  If instead a thick flux rope of radius $\sim$30$\arcsec$ (from the size of the overlying loops) is assumed, then the growth time becomes $\sim$90 s.  The observations show the growth time of 
the formation of blob `B' is about 2 min (at $\sim$07:28 UT initiation starts and at 
$\sim$07:30 UT disconnection of blob `B'). Therefore, the observational growth time for 
 ballooning instability (i.e. blob formation) is close to the theoretically predicted time if thicker flux rope of radius $\sim$30$\arcsec$ is assumed. It is difficult to reconcile the estimates for the onset condition and the growth time with the observations using the same value for the thickness of the flux rope.

  Single hard X-ray source seen in the AIA 1600 \AA \ image at 07:26:17 (overlaid by RHESSI contours) 
  was associated with the pre-heating or formation of helical current sheet below the prominence (shown in AIA 171 and 131 \AA). Moreover, we observe a hot flare loop below the prominence at 07:26:33 UT in AIA 131 \AA.
   The temperature enhances slowly due to plasma heating. This heating most likely increase 
  the plasma $\beta$, which supports the onset of ballooning instability. A blob shaped structure is formed at the top 
  (blob `A') possibly due to the ballooning instability \citep{shibasaki1999,shibasaki2001}. Later on, flare progresses and bright 
  loop was seen in 131 \AA \ images before the generation of another blob `B'. The another blob `B'
   was possibly also generated by the ballooning instability as a consequence of the heating seen in the soft X-ray flare. The magnetic 
   reconnection takes place at the time of leg detachment and the eruption of blob `B' (like a plasmoid eruption) and most of the non-thermal particles
    are accelerated starting at this time, as shown by footpoint sources. The eruption of blob `B' and hard X-ray
    profile peaks are well correlated. Additionally, it should be noted that the instability occurred in the northern leg (formation of blob `B') of the 
    flux rope. This may be due to the thickness of the leg. The northern leg is thinner in comparison to the southern one. According to theory of ballooning instability, thinner structures are more unstable in comparison to the thicker one under the coronal conditions.
 The above estimates show that the ballooning instability might have triggered in the considered event if preheating due to flare precursors raised the plasma beta and if the northern leg of the prominence was contained in a very thin flux rope. In this case, the blobs `A' and `B' could be produced by ballooning rather than by a helical kink. In principle, their shape is consistent with either instability. Although a much thicker flux rope appears far more likely, so that ballooning in the present event is not very likely, let us nevertheless discuss the possible role of the ballooning instability in triggering eruptions in more general terms. When preheating raises the plasma pressure sufficiently for the instability to occur, the magnetic flux can initially be changed by only a small amount. This is due to the small value of plasma beta in the corona. The instability is driven by the pressure, with part of the released thermal energy turning into the kinetic energy of the ballooning section(s) of the flux rope and the other part raising the magnetic energy. For $\beta< 1$, the magnetic energy can increase by only a small amount. Nevertheless, this small change could push the configuration to a point of magnetic instability. The resulting release of magnetic energy must be much larger than the initial minor increase, because a considerable (not a minor) part of the stored magnetic energy is required to power a CME and large flare  (e.g., \citealt{emslie2004,emslie2005}). This release
of magnetic energy should then be considered the main trigger of the eruption. Therefore, the ballooning instability can serve
only as a minor trigger or ``final drop" that leads to the eruption.

\subsection{Estimations for the Torus Instability}
\begin{figure*}
\centerline{
\includegraphics[width=10cm]{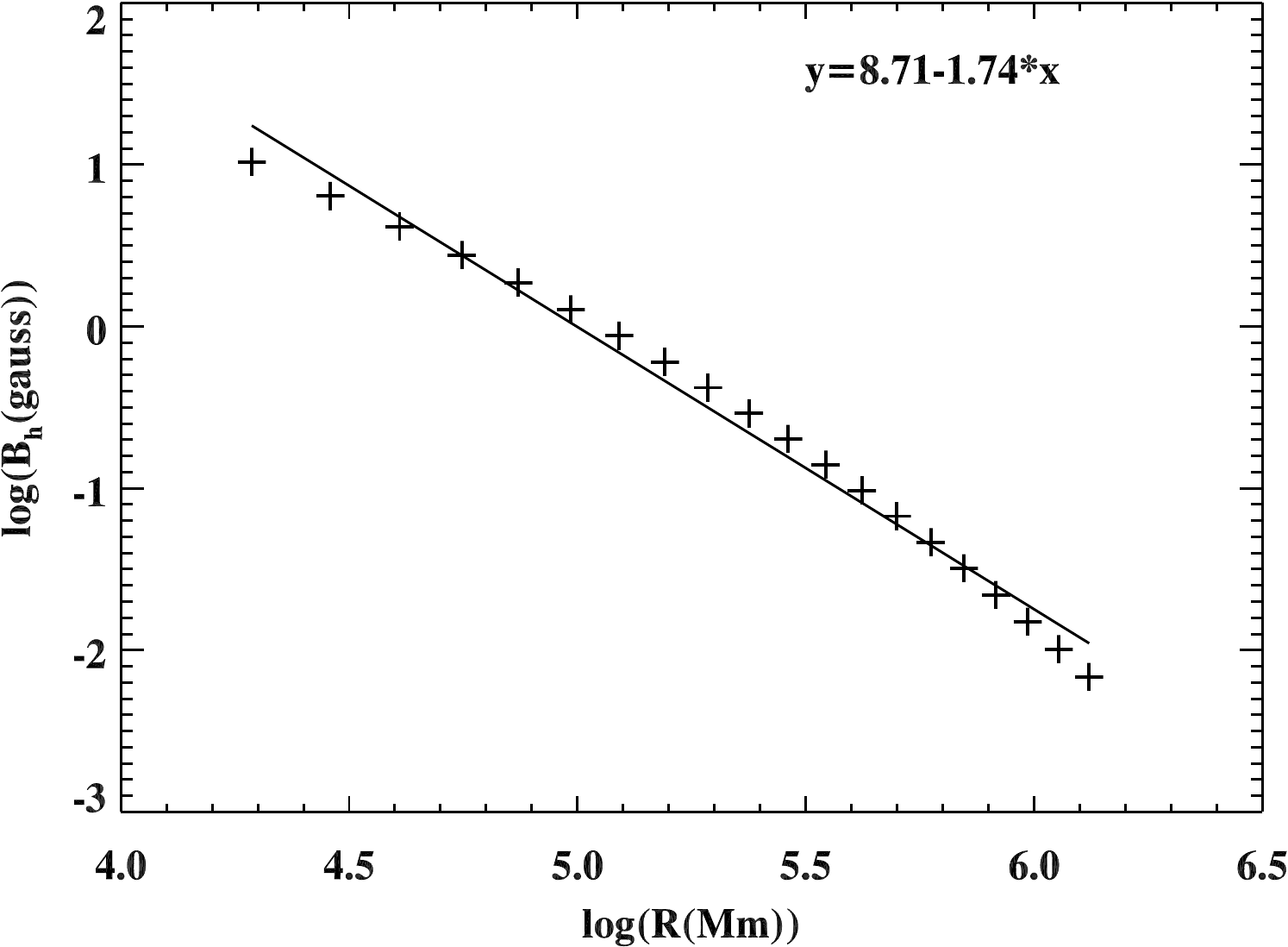}
}
\caption{The horizontal magnetic field strength (B$_h$) vs. height (R) plot (in logarithm) obtained from PFSS model. The symbol of `+' represents calculated data whereas the solid lines indicate a linear fitting to the data points. The slope gives the value of
the decay index. This plot shows a fitting for 0.10 R$_\odot$ to 0.65 R$_\odot$) heights from the solar surface.}
\label{torus}
\end{figure*}

 Finally, we consider the torus instability. Since the helical kink instability tends to saturate relatively quickly, the torus instability is a candidate for the acceleration of the CME in the later phase of the eruption. However, it may also be triggered much earlier, as an alternative to helical kink or as a process acting simultaneously. We note that both blobs `A' and `B' were formed significantly before the fast rise of the soft and hard X-ray emissions. It is possible that the kink instability formed the blobs, beginning around 07:25 UT, but the onset of the torus instability led to their main acceleration, along with the main flare emissions, after 07:29 UT. 
In this section,
we check the threshold criteria of the torus instability in the present event as given in eq. (1).
For this purpose, the overlying magnetic field was estimated from the observed magnetic field over the
solar surface based on PFSS model \citep{alt1969,sch1969,hoe1982,wang1992}. In this model,
it is assumed that the magnetic field is potential everywhere between the photosphere and a
spherical source surface. The magnetic field data at solar
surface was taken from SDO/HMI magnetogram \citep{graham2003}. At different heights from the solar surface, the modulus of horizontal component of magnetic field was averaged over an area of the active region where the eruption took place. Therefore, 
the profile of horizontal magnetic field strength (B$_h$) as a function of height (R) was obtained.
Figure \ref{torus} shows the horizontal magnetic field strength vs. height to estimate the decay index. We choose the heights of 0.10 R$_\odot$ to 0.65 R$_\odot$ above the solar surface, because the acceleration of the unstable flux rope started at these heights (refer to Figure \ref{ht}). The field strength and height are in logarithm. Thus, the decay index (n) is the slope of
a linear fitting to the data points, as shown in the Figure \ref{torus}. This average decay index for the above mentioned height range was obtained as 1.74. The decay index decreases downward, but only weakly, so that it exceeds the threshold of the torus instability in the whole considered range. Hence, torus instability can be the cause of the main acceleration of the unstable flux rope. Unlike the ballooning instability, it acts on the flux loop as
a whole, displacing it as a whole, and it 
operates under the low-beta conditions relevant in active regions.

\section{RESULTS AND DISCUSSION}

We present high resolution multiwavelength observations of the helical kink instability generated in the flux rope supporting a prominence , which caused a CME and associated M-class flare on 24 February 2011. Initial brightening below the prominence suggests the heating and formation of current sheet due to helical kink instability, associated with magnetic reconnection. The twist may be stored in the flux rope due to twisting and shearing motions at the footpoints. The addition of twisted flux to the rope by reconnection with the ambient field (if the added flux
makes an angle with the axis of the existing rope, so that it winds about the axis after the
reconnection) is a further possibility to reach the condition for the onset of the instability. The indication for a helical brightening observed prior to the eruption supports  this scenario.
    The formation of blobs `A' and `B' jointly is the strong indication of helical kink instability in the flux rope.   
The development and eruption of blob `B' plays an important role for initiating the magnetic reconnection and associated particle acceleration. The non-thermal particle acceleration correlates well with the acceleration of blobs `A' and `B' at the flux rope apex, which agrees the classical picture of eruptive flares.
In RHESSI reconstructed images, we observe the footpoint sources at the both side of polarity inversion line at the ends of the kinked flux rope. Ribbons flux enhances after the detachment of northern leg of the flux rope, due to particle acceleration and it matches well with the hard X-ray flux profile. 
 
According to \citet{gilbert2007}, the flux rope model is naturally applicable to the kinking phenomenon. The magnetic topology leading to a full, partial, and failed eruption can be understood by considering where reconnection occurs with respect to the prominence. The location of reconnection and the formation of X-type neutral lines in the two-dimensional flux rope model can occur completely below the prominence and its supporting flux rope, resulting in a full eruption. Reconnection can also occur completely above the prominence (i.e., a failed eruption) or within the flux rope itself, resulting in a partial filament or partial cavity eruption \citep{gilbert2001,gilbert2007}. In our case, reconnection probably occur below the blob `B' of the flux rope, which results in detachment of its leg generating the CME and particle precipitation along the footpoints of the flux rope (forming two hard X-ray footpoint sources). This scenerio is in agreement with the standard flare model i.e. CSHKP \citep{carm1964,stur1966,hirayama1974,kopp1976}.

We found another interesting feature in the CME speed, which showed the oscillatory pattern during 07:27-07:34 UT. Subsequently, the CME leaves the AIA field of view. Recently \citet{sham2010} also reported the quasi-periodic oscillation using SOHO/LASCO C2 and C3 data. The periods of quasi-periodic oscillations was found in the range 48--240 minutes, tending to increase with height.  They expect the morphology of the slowly erupting CME can be related to flux rope configuration. \citet{krall2001} suggested that when the geometry of the flux rope changes continuously, the oscillations are likely to change in character as the flux rope expands. In agreement to this, the periods of the oscillations of CMEs tend to increase as the distance increases. In our case, we observe oscillations with a period of $\sim$2 minute in the low corona, which additionally support the hypothesis that the CME was triggered in a flux rope structure. 

Several authors have shown that a kinking and rising flux rope triggers the formation of a helical current sheet in addition to the formation of the vertical (flare) current sheet (e.g., \citealt{gerrard2001}, \citealt{kliem2004}, \citealt{kliem2010}). The helical current sheet wraps around the legs and passes over the upper section of the flux rope in the interface to the ambient field.  The present event shows initial brightenings and indications of a bright helical structure below the apex of the flux rope in blobs `A' and `B', which possibly may be associated with a helical current sheet.

This event also exhibits features which are consistent with the occurrence of ballooning instability.
According to \citet{st2008}, at the nonlinear stage of its development the ballooning instability can result in the reconnection of magnetic field lines and in the flare energy release, thus playing part of a trigger for solar flares. The development of a plasma ``tongue", initiating the reconnection process, naturally explains the formation and ``tearing-off" of plasmoids, which exactly determine destabilization and rising of overlying magnetic structures. Within the framework of ideal magnetohydrodynamics the excitation of the ballooning instability in a toroidal coronal loop with a radius of cross section `r' and a radius of curvature `R', the ballooning instability can be excited by kink oscillations when the plasma beta parameter $\beta>2r/R$ \citep{tsap2008}. Preheating prior to the impulsive phase of the flare generally increases the plasma beta. \citet{tsap2008} also mentioned that the excitation of the ballooning instability may happen more than once and can cause several blob-like sources and their detachment owing to magnetic reconnection. In the present event, the formation of multiple blobs `A' and `B', associated with reconnection, was seen. Thus, as an alternative to kink instability, the ballooning instability may have formed the blobs `A' and `B', which played an essential role in the dynamics of the event. 

 However, our quantitative estimates in Section 3.2 shows that the onset condition of this instability could only be satisfied if the northern leg of the erupting prominence was held against gravity by a flux rope of very small cross-section (minor radius). The fact that the overlying loops moved synchronously with the rising prominence suggests that the flux rope did in fact have a much larger cross-section, extending up to, or nearly up to the overlying loops. Moreover, it is unlikely that the cross-section of the flux rope differs strongly between the two legs, because external forces that could squeeze one leg are not available in the low beta corona. Finally, the morphology of blobs `A' and `B' is consistent with the occurrence of kinking but not with the occurrence of ballooning. The kink mode moves the magnetic axis in the center of the flux rope, while the ballooning mode operates on the surface of the flux rope. Both loops that developed into blobs `A' and `B' are suggestive of changes in the center of the flux rope. In summary, although several aspects of the eruption are consistent with ballooning, other aspects argue against it, so that the occurrence of this instability is not very likely. Therefore, it is more likely that blobs `A' and `B' were formed by the helical kink instability.

It is found that the coronal field above the prominence has a decay index which exceeds the threshold of torus instability in the whole height range of strong acceleration of blobs `A' and `B' (above $\sim$0.1 R$_\odot$). Therefore, this instability can have been the main driver of the CME and associated flare from the onset of the strong acceleration at $\sim$07:29 UT.

In conclusion, we have shown multiwavelength observations of the helical kink instability initiating a CME and the associated particle acceleration using high resolution data from SDO/AIA combined with RHESSI. This event provides a strong evidence of helical kink instability as a triggering mechanism for the CME and of the torus instability as a CME driver from the onset of the main acceleration. Considering that the energy for the eruption must mainly be stored in the magnetic field, the
ballooning instability seems to be less capable to drive a CME. However, although not very likely, the
formation of two localized, semicircular plasma blobs before the main acceleration of the CME may have been due to this instability.
Different from the earlier cases with the involvement of the helical kink instability \citep{ji2003,alex2006,cho2009}, the present event does not show any evidence for leg-leg reconnection, which may have  accelerated the particles along the flux rope legs in those events. Future study of similar events using high resolution data may shed more light on the cause of  plasma instabilities and their relations with the eruption processes.

\acknowledgments
We express our gratitude to the referee for his/her valuable suggestions and helping in the interpretation of the observations which improved the manuscript considerably. SDO is a mission for NASA’s Living With a Star (LWS) program. RHESSI is a NASA Small Explorer. PK thanks to Prof. P. K. Manoharan, Prof. Kwang-Son Choe and Prof. K. Shibasaki for the fruitful discussions. We acknowledge SOHO, RHESSI and BIRS for providing the data used in this study. SOHO is a project of international cooperation between ESA and NASA.   
\bibliographystyle{apj}
\bibliography{references}
\clearpage


\end{document}